\documentclass{aa}

\usepackage{graphicx}
\usepackage{tabularx}
\usepackage{txfonts}
\usepackage{placeins}

\newcommand{\Ha}{\mbox{\ensuremath{\mathrm{H}\alpha}}}

\newcommand{\hii}{\ion{H}{II}}

\begin{document}

   \title{Method on using shadow altitude to remove geocoronal \Ha}

   \author{Wai-Kiu Ricky Wong\inst{1,2}\thanks{Email: wkrickywong@link.cuhk.edu.hk}
        \and Renbin Yan\inst{1,2,3}\thanks{Email: rbyan@cuhk.edu.hk}
        \and Zesen Lin\inst{4,1,3}
        }

   \institute{Department of Physics, The Chinese University of Hong             Kong, Shatin, N.T., Hong Kong, China\\
   \and
   JC STEM Lab of Astronomical Instrumentation, The Chinese University of Hong Kong, Shatin, N.T., Hong Kong, China\\
   \and CUHK Shenzhen Research Institute, No.10, 2nd Yuexing Road, Nanshan, Shenzhen, China\\
   \and
   Institute for Astrophysics, School of Physics, Zhengzhou University, Zhengzhou, 450001, China
        }

   \date{}

  \abstract{Spectroscopic surveys allow spatially resolved spectroscopy of galaxies to study their interstellar medium (ISM). However, observations of Galactic \Ha\ emission are contaminated by geocoronal \Ha\ emission. The latter is known to depend on the shadow altitude, a geometric parameter relating the line of sight to Earth's shadow cone. Using fibres on blank skys from the MaNGA Stellar Library (MaStar) project of the Sloan Digital Sky Survey IV (SDSS-IV), we establish an empirical relation between the geocoronal \Ha\ emission and the shadow altitude, with a root mean square fractional scatter of 24.41\%. This relation can be used to predict geocoronal \Ha\ emission so that it can be removed from observed spectra. This removal method is advantageous when the observed targets are extensive in the sky, and it does not require a large velocity separation between the observed target and the local standard of rest. This will enable reliable studies of Galactic \Ha\ in intermediate spectral resolution integral field spectroscopic surveys. We also find tentative evidences for the dependences of geocoronal emission on solar activity and the distance between the Earth and the Sun.}


   \maketitle
   \nolinenumbers

\section{Introduction} \label{sec:intro}
Direct observation of \Ha\ emission from extended sources in the local Universe often proves difficult due to contamination from geocoronal \Ha\ emission in the sky. Usually, sky spectrum subtraction on small angular-sized objects, such as extragalactic objects or stars, is done by subtracting the pure sky spectrum observed at nearby pointings, which would contain a similar level of geocoronal \Ha. Unfortunately, when the target is spatially-extended and is larger in angular size than the scale on which geocoronal \Ha\ varies, the sky spectrum observed at a separated sky location would not be sufficient for subtracting geocoronal \Ha\ cleanly.

\Ha\ emission from ionised interstellar medium (ISM) in the Milky Way is one prime example of a spatially extended source impacted by the sky subtraction challenge mentioned above. \Ha\ emission line is widely used as a tracer for the active star-forming (SF) region in galaxies, as young massive stars produce large amounts of ionising photons, ionising surrounding clouds. \Ha\ emission from diffuse ionised gas (DIG) is also important since it contributes to 30\,\% to 60\,\% of the total \Ha\ flux we observed from galaxies \citep{2007ApJ...661..801O, 1996AJ....111.2265F, 1996AJ....112.1429H, walterbos_1998}, which will greatly bias the metallicity measurement of the \hii\ regions in SF galaxies as DIG introduces significant bias to the observed line flux \citep{zhang2017sdss}.

In order to study the interaction of \hii\ regions and surrounding DIG in greater detail, a high spatial resolution of parsec scale will be needed to resolve the small-scale structure of the ISM. However, common spatially-resolved spectroscopic surveys on galaxies such as the Mapping Nearby Galaxies at Apache Point Observatory (MaNGA) project (on $\sim$ 1-2 kpc scale, \citealp{2015ApJ...798....7B}) or PHANGS-MUSE (on $\sim$ 50 pc scale, \citealp{phangs-muse}) cannot provide us with the high spatial resolution required. Surveys like SDSS-V Local Volume Mapper (LVM) \citep{2019BAAS...51g.274K, drory2024sdss} and Affordable Multiple Aperture Spectroscopy Explorer (AMASE) \citep{Yan_2020, Yan2024}  will be able achieve this. Before their data are available, one may be able to use background fibres from surveys like MaNGA or MaNGA Stellar Library (MaStar) to study DIG. However, we have to find a way to subtract geocoronal emission cleanly.

If the spectral resolution is sufficiently high, the velocity separation (VLSR) between the local standard of rest (LSR) frame and the local rest frame of the observatory can be used to separate the geocoronal \Ha\ (${\rm \sim 7\ km\ s^{-1}}$ FWHM) from the Galactic \Ha\ (${\rm \sim 20\ km\ s^{-1}}$ FWHM) \citep{Haffner_2003, mierkiewicz2006geocoronal} by simple double Gaussian fit. Fabry-Perot interferometers have long been used to observe \Ha\ emission from the Galactic ISM, given their high spectral resolution. Some examples include the Wisconsin \Ha\ Mapper Northern Sky Survey (WHAM-NSS) installed at Kitt Peak \citep{Haffner_2003} (R$\sim$ 25000, ${\rm 12\ km\ s^{-1}}$), the Wisconsin \Ha\ Mapper Southern Sky Survey (WHAM-SSS) installed at Cerro Tololo \citep{2010AAS...21541528H} (R$\sim$ 25000, ${\rm 12\ km\ s^{-1}}$ ), and the Dual Etalon Fabry-Perot Optical Spectrometer (DEFPOS) installed at T\"UB\'ITAK National Observatory (TUG) \citep{M.Sahan_2005} (R$\sim$ 11000, ${\rm 27\ km\ s^{-1}}$). In WHAM, this method worked well because the VLSR is usually offset by ${\rm \sim 18 - 42\ km\ s^{-1}}$ from solar peculiar motion and Earth's orbital motion, which provides sufficient separation of two emission lines for their spectral resolution. However, the spatial resolution of a typical Fabry-Perot Spectrometer is poor, for example, 1 ${\rm deg^2}$ for WHAM, rendering them unsuitable for the high spatial resolution required to study any internal structure of ISM.

Others have tackled this issue by using the WHAM data as a constraint on their flux calibration \citep{SHASSA_2001, finkbeiner2003full}. Newer technique independent of WHAM data uses the flux ratio between geocoronal \Ha\ emission and OH ${\rm \lambda 6554}$ sky line emission, and its relation to the solar altitude has been used \citep{Zhang_2021}, however, a need for large VLSR ${\rm >\ 25\ km\ s^{-1}}$ is used in their fitting given their limited data set consists of both Galactic \Ha\ and geocoronal \Ha.

We develop a method that can reliably estimate the geocoronal \Ha\ flux from the spectrum based on the shadow altitude with the much larger data set from MaStar, a spectroscopic survey for stars. It does not rely on the large VLSR between geocoronal \Ha\ and Galactic \Ha\, and can work on a relatively low spectral resolution. This method can be used to subtract geocoronal \Ha\ contamination to reveal the underlying Galactic \Ha.

This paper is structured as follows, we describe the dataset and data preprocessing in Sect.  \ref{sec:data_description}, the sky background model in Sect. \ref{sec:skybg_model}, fitting of the spectrum in Sect. \ref{sec:spec_fitting}, the result in Sect. \ref{sec:results}, and discussion in Sect. \ref{sec:discussion}.
\section{Data description} \label{sec:data_description}
\subsection{MaStar} \label{subsec:mastar}
MaNGA Stellar Library (MaStar) project \citep{2019ApJ...883..175Y} is one of the projects of the fourth generation of the Sloan Digital Sky Survey (SDSS-IV) \citep{SDSS-IV_Blanton}, which uses the integral field unit (IFU) \citep{MaNGA_IFU_15} from MaNGA \citep{2015ApJ...798....7B} to obtain the spectra of stars with a spectral resolution of R${\rm \sim 1800}$ during bright time.

The star located at the centre of each IFU can only contaminate a finite angular area, hence the fibres at the periphery of the IFU pointing at the background can be used to study the Galactic ISM, and their spectra will be the data used in this work.

However, directly using these MaStar data to obtain the Galactic spectra comes with a few challenges. Since the observations were done during bright times, significant sky background contamination was present in our data. Although the MaStar data reduction pipeline did handle the sky background subtraction, it was not designed for measuring the Galactic gas emission, as these were not the original targets planned for MaStar. The sky fibres targeting empty sky around the IFUs would contain different levels of the Galactic gas emission and geocoronal \Ha\ to the IFUs due to spatial variation. This is especially true when observing extended targets such as the \hii\ regions and DIG within the Milky Way. Therefore, we shall construct a model of the sky background first in order to separate the Galactic \Ha\ from the geocoronal \Ha, the sky background model is explained in detail in Sect.  \ref{sec:skybg_model}.

\subsection{Data selection} \label{subsec:data_selection}
First, all MaStar visits of all stars are chosen, where a visit is a single night of observation for a star, and a star may have one or more visits. Quality flags tabulated in Table \ref{table:visit_bitmask} are used to remove any visits that will affect our peripheral fibres, we consider the filtered visits to be good visits in this study. Moreover, small 7-fibre mini-IFUs pointing at standard stars for calibration purposes are not included, given that the starlight could easily contaminate all of the 7 fibres.

\begin{table}[htbp]
    \caption{Bit name, bit number, and description of the quality flags used to remove poor quality visits for the peripheral fibres.}
    \label{table:visit_bitmask}
    \centering
    \small
    \begin{tabularx}{\columnwidth}{l c X}
        \hline\hline
        Bit name & Bit number & Description \\
        \hline
        \texttt{HIGHSCAT}   & 2  & High scattered light in one or more frames. \\
        \texttt{BADFLUX}    & 3  & Bad flux calibration. \\
        \texttt{POORCAL}    & 5  & Poor throughput. \\
        \texttt{SEVEREBT}   & 12 & Severe blowtorch artifact. \\
        \texttt{NONSTELLAR} & 16 & Non-stellar spectrum based on visual inspection. \\
        \texttt{CRITICAL}   & 30 & Critical failure in one or more frames. \\
        \hline
    \end{tabularx}
\end{table}

\begin{figure*}[ht]
\sidecaption
    \includegraphics[width=12cm]{{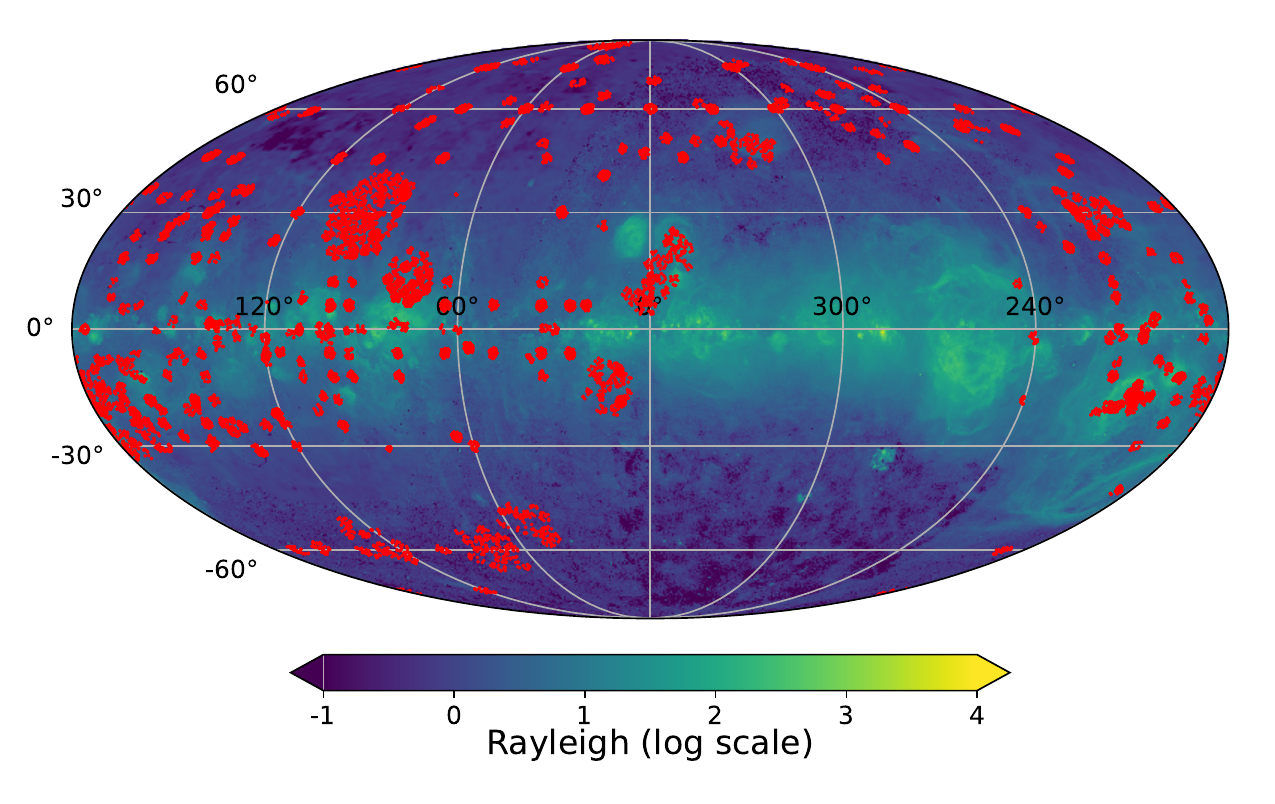}}
    \caption{Full sky Galactic \Ha\ map from \citet{finkbeiner2003full}, with the MaStar visits filtered for quality flags in Table \ref{table:visit_bitmask} overplotted (red crosses).}
    \label{fig:Halpha_full_sky_map}
\end{figure*}

Then, we use the full sky Galactic \Ha\ map from \citet{finkbeiner2003full}. This map combines observational data from WHAM \citep{Haffner_2003}, the Virginia Tech Spectral-Line Survey (VTSS) \citep{VTSS_1998, VTSS_1999}, and the Southern H-Alpha Sky Survey Atlas (SHASSA) \citep{SHASSA_2001} together to cover the entire night sky. The WHAM provided a high spectral resolution (${\rm R \sim 25000}$, ${\rm 12\ km\ s^{-1}}$) in 1 ${\rm deg^2}$ spatial resolution, while both VTSS and SHASSA are narrow band observations with arc-minute scale spatial resolution (1.6$\arcmin$ pixel width for VTSS, 0.8$\arcmin$ pixel width for SHASSA), with their flux calibrated by WHAM observation. We can treat the full sky Galactic \Ha\ map as the ground truth, but with a lower spatial resolution than the MaStar IFU. The full sky Galactic \Ha\ map with the location of each visit filtered for the quality flag in Table \ref{table:visit_bitmask} over-plotted is shown in Fig. \ref{fig:Halpha_full_sky_map}.

\begin{figure}[ht]
\begin{center}
    \includegraphics[width=0.90\linewidth]{{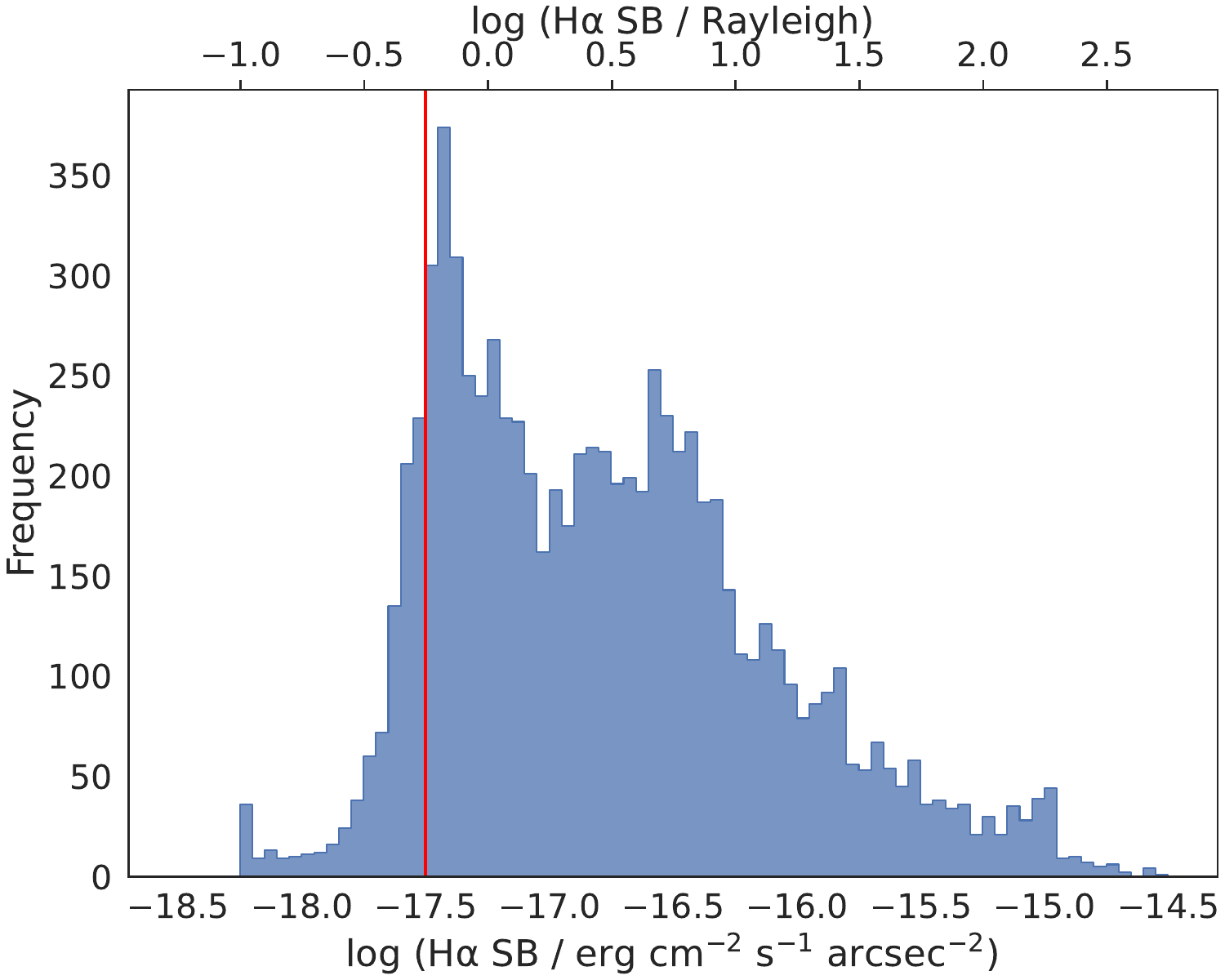}}
    \caption{Histogram of Galactic \Ha\ SB retrieved from the full sky \Ha\ map using the unique pointing of each IFU, with the criterion used overplotted as the red vertical line.}
    \label{fig:Halpha_flux_hist}
\end{center}
\end{figure}

We utilise this full sky \Ha\ map to separate the MaStar pointings at low or high Galactic \Ha\ locations in the following way. For every MaStar pointing with good visits, we compare its pointing to the Finkbeiner full sky \Ha\ map for the expected Galactic \Ha\ surface brightness (SB), then we separate the IFUs into their corresponding group based on the \Ha\ SB threshold, which is chosen at ${\rm 10^{-17.5}\,erg\,cm^{-2}\,s^{-1}\,arcsec^{-2}}$. This value is near the peak of the distribution of \Ha\ SB (see Fig. \ref{fig:Halpha_flux_hist}) while providing enough IFUs for the low Galactic \Ha\ group for statistical studies. Notice this threshold value is arbitrary and can be chosen as any other reasonable value, this will be further discussed in Sect.  \ref{subsec:intrinsic_scatter_sources}.
The distribution of the Galactic \Ha\ SB extracted from the full sky \Ha\ map of each unique pointing is shown in Fig. \ref{fig:Halpha_flux_hist}. The low Galactic \Ha\ pointings are assumed to have no or little Galactic \Ha\ SB and are used to calibrate our sky background model, including the geocoronal \Ha\ flux. The high Galactic \Ha\ IFUs can be used to study the Galactic \Ha\ emission for later science purposes.

We use the version of the \texttt{mgCFrame} file with linear wavelength grid (the \texttt{LIN} version) provided by the MaStar data reduction pipeline\footnote{\url{https://www.sdss4.org/dr17/manga/manga-data/data-model/}}. This is a data product of the pipeline containing the sky-subtracted and flux-calibrated spectrum for each fibre. However, as mentioned, the sky subtraction procedure of the standard pipeline would subtract away both the Galactic line emission and the geocoronal line emission, we add back the subtracted sky model to each spectrum for the subsequent analysis. There is one \texttt{mgCFrame} file per exposure. Any exposure with an exposure time of less than $\rm{890\,s}$ and a transparency of less than 0.8 is excluded in our analysis due to their potential low signal-to-noise ratio (S/N). We also exclude astronomical twilight exposures to prevent significant sunlight scattering and refraction, which is difficult to subtract correctly.

The wavelength window is limited to be from 6530\,\AA\ to 6600\,\AA\ for computational speed, which includes the \Ha\ line, $\mathrm{[NII] \lambda \lambda 6550,6585}$ doublet lines and eight skylines.

In total, we have 7299 exposures with 842 unique pointings in the low Galactic \Ha\ group and 30277 exposures with 2776 unique pointings in the high Galactic \Ha\ group after all the data preprocessing steps and spectrum fittings described in this section, and Sects. \ref{subsec:fiber_exclusion}, \ref{subsec:stacked_spec}, \ref{sec:skybg_model}, and \ref{sec:spec_fitting}.

\subsection{Fibre exclusion} \label{subsec:fiber_exclusion}
The fibres containing stellar light within each IFU are first excluded since the star is not our target and is a source of contamination. We simply exclude the fibres if the median science flux, $f_{m}$, of that fibre within a wavelength window without any Galactic emission line (6450~\AA\ - 6500~\AA) is larger than a flux threshold.

\begin{figure}[htbp]
\begin{center}
    \includegraphics[width=0.90\linewidth]{{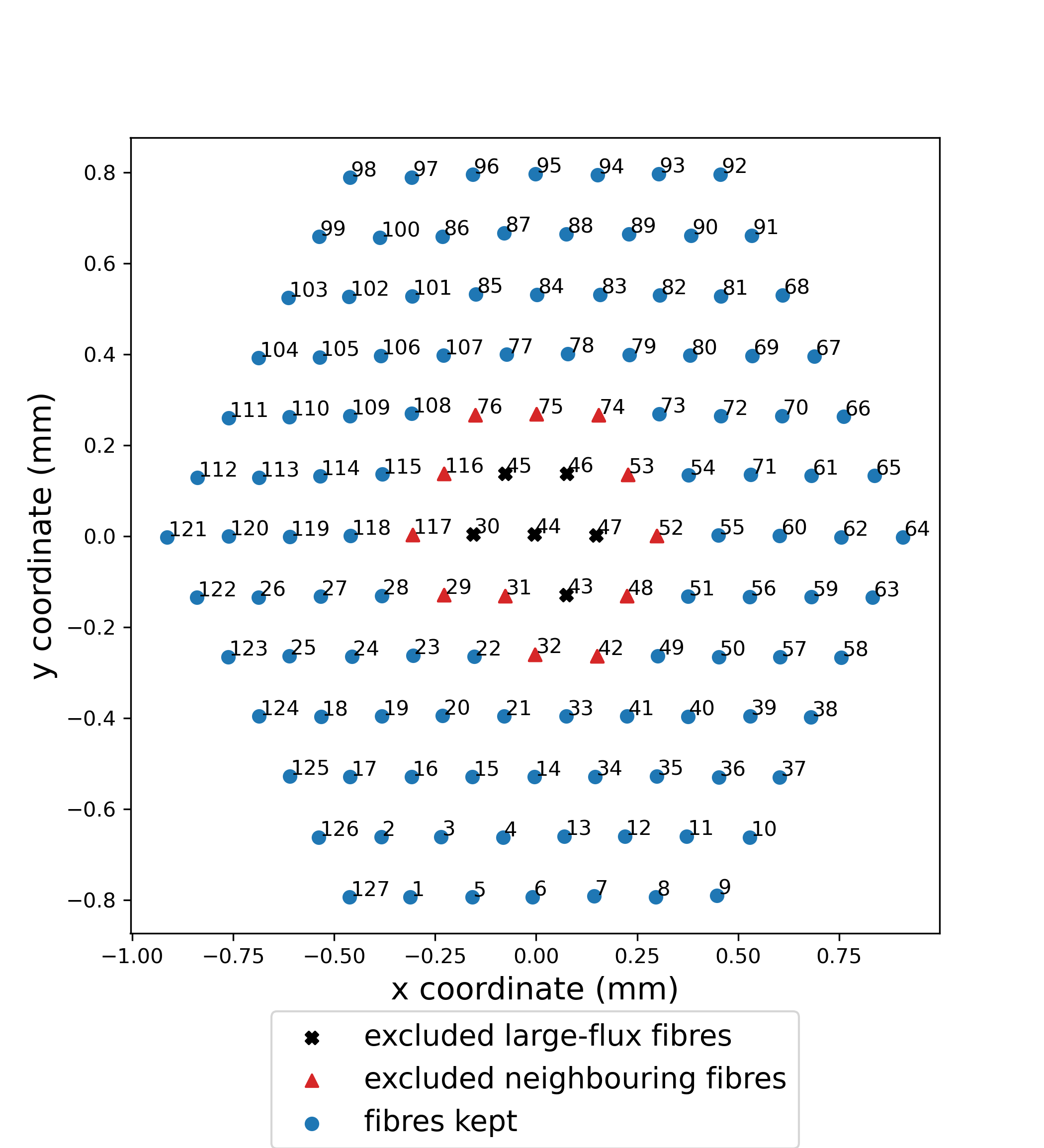}}
    \caption{Hexagonal fibre arrangement within an example 127 fibres IFU (plate 8051, modified Julian date 57057, mangaid 27-1858, exposure id 193985, IFU-design 12705). The excluded large-flux fibres are marked with black dots, their neighbouring fibres are marked with red dots, and the remaining fibres are marked with blue dots. The numbers above the dots are fibre ids. The excluded large-flux fibres are located at the centre as expected for the star location.}
    \label{fig:fiber_exclusion}
\end{center}
\end{figure}

The flux threshold is determined by sky fibres, which are single fibres targeting empty sky regions around the IFUs. We first calculate the median flux of each sky fibre within the wavelength window in the same way as we do for science IFUs. Then, for each exposure of each plate, we compute the median of these median fluxes among all sky fibres on that plate, $m$, and their standard deviation, $s$. We define $m+2s$ to be the flux threshold to exclude fibres with potential stellar light contamination, which obviously will have larger median flux than that of the sky fibres. To take into account the effect of starlight leaking into nearby fibres, we also exclude any neighbouring fibres of fibres with large median flux. An example of this fibre exclusion step is illustrated in Fig. \ref{fig:fiber_exclusion}.

IFUs pointing at standard stars for calibration purposes are not included since there are only 7 fibres within these IFUs, and the star would contaminate the entire IFU.

\subsection{Stacked spectrum}
\label{subsec:stacked_spec}
To ensure a high S/N, the spectra within each IFU are stacked together before any fitting. We take the mean spectrum among all the remaining fibres of each IFU for each exposure. Thus, we effectively have a single spectrum with a larger field of view (FOV). Indeed, this stacking will lower our effective spatial resolution. For MaStar IFU, the FOV is from 12$\arcsec$ to 32$\arcsec$ (${\rm \sim 0.12\ pc\ -\ 0.31\  pc\ at\ 2\ kpc}$) depending on the size of the IFU, but it still yields higher spatial resolution compared to the Finkbeiner full sky \Ha\ map.

\section{Sky background model} \label{sec:skybg_model}
The \texttt{mgCFrame} files contain sky background and must be subtracted away before fitting for Galactic lines at the high Galactic \Ha\ plates. The sky background is modelled in three major components as shown below,

\begin{equation} \label{eq:skybg_model}
\begin{aligned}
    &q(\lambda)=\\
    &af(\lambda - k_f) + b + \sum_{i}A_{i, \mathrm{sky}}\exp(-\frac{(\lambda - (\lambda_{i, \mathrm{sky}} + k_{\mathrm{sky}}))^2}{2\sigma_i^2}), \\
\end{aligned}
\end{equation}
where each term is labelled as follows.
\begin{itemize}
    \item The first term represents those components in the sky background that are due to scattered sunlight by the moon and the atmosphere, where $a$ is the amplitude, $f(\lambda - k_f)$ is the shifted solar template, and $k_f$ is the shifting in wavelength of the solar template.
    \item The second term, $b$, is the remaining sky background continuum after excluding any components with the solar spectrum.
    \item The third term is the skylines, where $A_{i, \mathrm{sky}}$ are the amplitude of the Gaussian peaks, $\lambda_{i, \mathrm{sky}}$ is the skyline centres, $k_{\mathrm{sky}}$ is the shifting in wavelength of the skylines, $\sigma_i$ is the post-pixelised instrumental line spread function (LSF) used as the width of the skylines, which varies slightly with wavelength, therefore $\sigma_i$ located at $\lambda_{i, \mathrm{sky}} + k_{\mathrm{sky}}$ is used.
    \item $q$ is the observed flux.
\end{itemize}

The details of each component are explained in the following sections.

\subsection{Solar template} \label{subsec:solar_template}
The majority of sky background contamination comes from the reflection of sunlight from the moon and scattering by the atmosphere, as the MaStar data were observed during bright time. Therefore, we use the observed solar spectrum from Jungfraujoch observatory \citep{1990apds.book.....D}, provided by the BASS 2000 website\footnote{\url{https://bass2000.obspm.fr/reference.html}{}} \citep{aboudarhambass2000} with a spectral resolution of 0.002\,\AA\ from 3000\,\AA\ to 10000\,\AA\ to be the template of our solar spectrum.

Since the template has a much higher spectral resolution and finer sampling than MaStar, we first apply a Gaussian convolution to approximate the instrumental effect, in which the mean of the standard deviation of the pre-pixellised MaStar LSF in the fitting wavelength window (6530\,\AA\ - 6600\,\AA) is adopted as the standard deviation of the kernel. Then a box kernel of 1\,\AA\ wide is used to lower the sampling to match the wavelength grid used in the \texttt{mgCFrame} file. The wavelength window used in these convolution steps is extended by 10\,\AA\ to avoid any boundary effect.

Moreover, the template provided is in air wavelength so it is converted into vacuum wavelength using the equation from Vienna Atomic Line Database 3 derived by N. Piskunov \citet{Ryabchikova_2015}. Notice the wavelength grid will not be uniform after this conversion, hence linear interpolation is used to enforce a uniform wavelength grid.

\begin{figure}[ht]
\begin{center}
    \includegraphics[width=0.90\linewidth]{{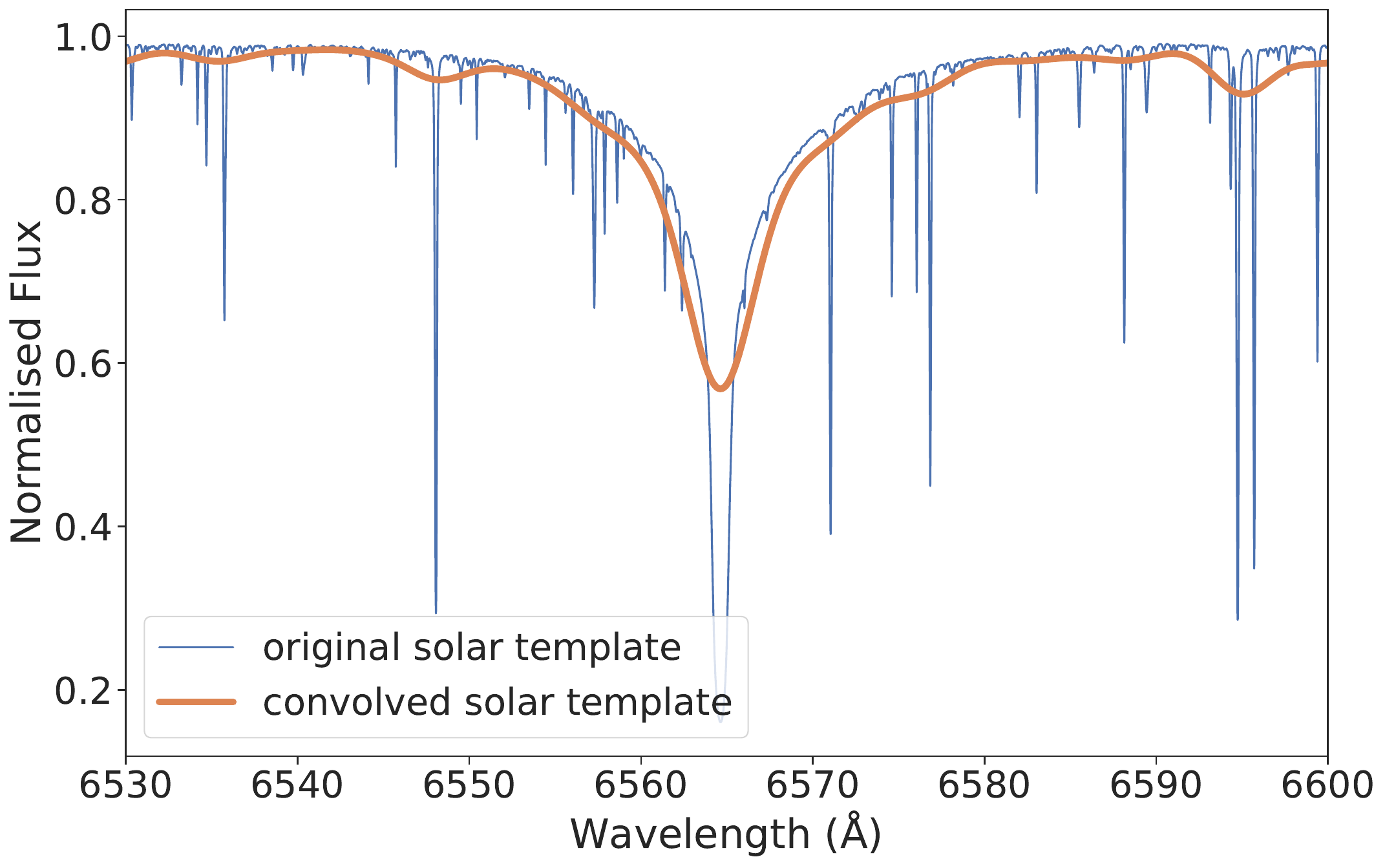}}
    \caption{The high spectral resolution solar template spectrum provided by BASS 2000 (blue curve) and the processed solar template spectrum in 1~\AA\ resolution (orange curve) are shown in vacuum wavelength and normalised flux, with no solar template shift ($ k_f = 0$).}
    \label{fig:solar_template}
\end{center}
\end{figure}

To remove any wavelength calibration issue, the wavelength of the solar template is allowed to shift by $k_f$, which is limited from -1\,\AA\ to 1\,\AA. At last, the processed solar spectrum template, $f(\lambda)$, with the same wavelength grid as the \texttt{mgCFrame} file is shown in Fig. \ref{fig:solar_template}.

The multiplicative free parameter, solar amplitude, $a$, is used to represent the strength of the solar template, which could be affected by the illumination fraction of the moon or the angular separation between the moon and the IFU.

\begin{figure}[ht]
\begin{center}
    \includegraphics[width=0.90\linewidth]{{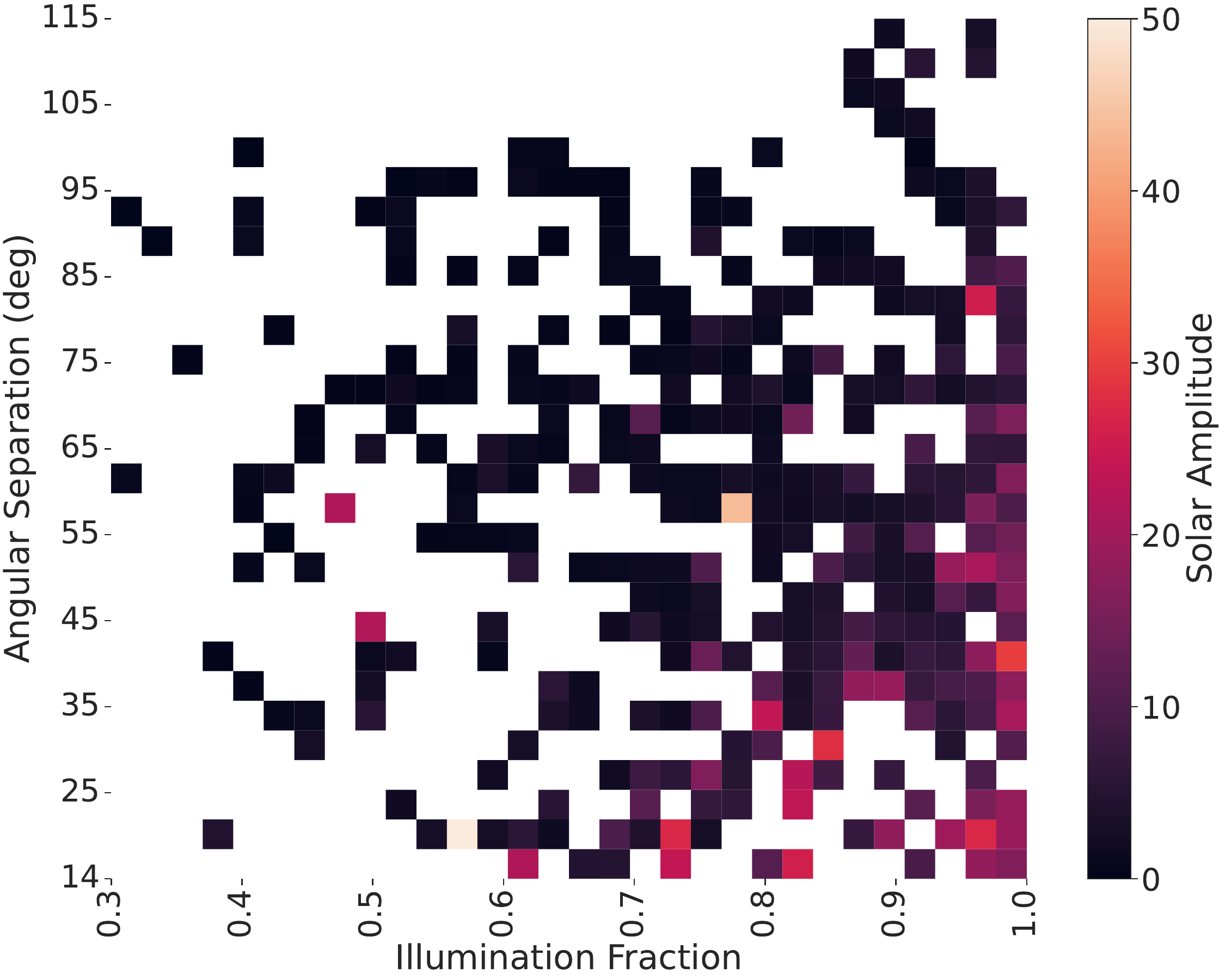}}
    \caption{Solar amplitude, $a$, at different angular separation between the moon and the IFU (y-axis) and linear illumination fraction (x-axis). Colour bar represents the median $a$ in each pixel. A clear trend of increasing $a$ towards the lower right corner can be seen.}
    \label{fig:moon_brightness}
\end{center}
\end{figure}

The distribution of solar amplitude, $a$, at different lunar angular separations and lunar illumination fractions is shown in Fig. \ref{fig:moon_brightness}. A clear trend of increasing $a$ towards the lower right corner can be seen, which corresponds to a small angular separation and large lunar illumination fraction, verifying the idea that the strength of the sky background is correlated to the reflected sunlight from the moon.

\subsection{Sky background continuum} \label{subsec:skycon}
The additive free parameter, $b$, is used to represent any sky background continuum other than the solar continuum. Given the small wavelength window we used, this sky background continuum contribution can be approximated as a flat contribution.

\subsection{Skylines} \label{subsec:skylines}
Skylines are extracted from the sky emission spectrum obtained by the Ultraviolet and Visual Echelle Spectrograph (UVES)\footnote{\url{https://www.eso.org/observing/dfo/quality/UVES/pipeline/sky_spectrum.html}}, a high-resolution optical spectrograph of the Very Large Telescope from the European Southern Observatory \citep{2003A&A...407.1157H}. Not all skylines measured in UVES are used, due to the lower spectral sampling of \texttt{mgCFrame} data. It is noted that the MaStar 90-th percentile FWHM is about 4\,\AA\ at \Ha, therefore, if two neighbouring skylines have their line centre difference < 1\,\AA, they are first combined as a single line, with their average wavelength to be the new line centre of the combined line. This is to reduce the number of skylines to be fitted and prevent degeneracy in the spectrum fitting. The fluxes of these two combined lines provided by UVES are simply summed together.

We further remove any skylines, combined or not, with a peak flux smaller than $2\times 10^{-17}\,\mathrm{erg\,cm^{-2}\,s^{-1}\,arcsec^{-2}}\,\text{\AA}^{-1}$ in the UVES observation, as there is a group of relatively weak skylines before combining the skylines (see Appendix \ref{sec:remove_weak_skylines}). The skylines provided by UVES are also in air wavelength; therefore, we converted them into vacuum wavelength following the same scheme as in Sect.  \ref{subsec:solar_template}.

The skyline components selected are located at 6534.85\,\AA, 6542.33\,\AA, 6545.84\,\AA, 6555.03\,\AA, 6570.60\,\AA, and 6579.11\,\AA, 6585.92\,\AA, 6598.46\,\AA, all of which except skyline 6585.92 \AA\ are OH skylines, and the geocoronal \Ha\ line at 6564.60 \AA\ (wavelength presented in vacuum wavelength). Each line is fitted with a single Gaussian. The centre of the Gaussian component of all skylines shares the same shifting, $k_{sky}$, with respect to the original wavelength centre, by 1\,\AA\ at most, to address any potential wavelength calibration problem in the \texttt{mgCFrame} file. The line width observed is dominated by the instrumental effect at our spectral resolution, hence we use the post-pixelised instrumental $\sigma_i$, as the width of the Gaussian, which varies slightly with wavelength.

\section{Spectrum fitting} \label{sec:spec_fitting}
\subsection{Low Galactic \Ha\ pointings} \label{subsec:low_gal_Ha}
The low Galactic \Ha\ IFUs are assumed to have no Galactic components, hence we can use Eq. \ref{eq:skybg_model} as the spectrum model, simply fitted by \texttt{optimize.curvefit} package from \texttt{scipy}. However, the fitting can still be improved by minimising the error introduced in the solar template, and we implemented an iterative refitting process to reduce the error, explained in the following section.

\subsection{Refit of solar template} \label{subsec:refit}
\subsubsection{Relative wavelength shift\\ between the solar template and skylines} \label{subsubsec:relative shift}
Both the solar template and skylines are allowed to shift freely in wavelength by 1\,\AA\ at most in the first fitting as both the solar template and MaStar data may have imperfect wavelength solutions. The median difference, $\Delta \lambda_{\rm median} = {\rm median}(k_{sky} - k_f)$, between their wavelength shift of all low Galactic \Ha\ IFU pointings is determined to be the systematic difference of the wavelength solution between the solar template and skylines. This median difference, $\Delta \lambda_{\rm median}$, is added as a constant shifting at the remaining refitting process.

\subsubsection{Refit of the flux in the solar template} \label{subsubsec:refit_flux}
The BASS 2000 solar template is from direct observation of the sun, which could be modified by the moon's reflection and the scattering of the atmosphere. Part of the fitting residuals due to imperfect template should scale linearly with the solar amplitude, $a$. Thus, we fit the residual as a linear function of $a$ with a Huber linear fit at each wavelength grid as shown below,

\begin{equation} \label{eq:huber_refit}
\begin{aligned}
    r(\lambda) = m(\lambda)a + c(\lambda),
\end{aligned}
\end{equation}
where $r(\lambda)$, $m(\lambda)$, $c(\lambda)$ are the fitting residual, slope, and y-intercept of the Huber fit at wavelength grid $\lambda$, respectively.

Huber fit is preferred over the usual linear fit to negate the effect of outliers having a disproportionally large impact on the resulting $m(\lambda)$ and $c(\lambda)$ \citep{huber2011robust}. The threshold used in Huber fit is 1.35.

$m(\lambda)$ and $c(\lambda)$ obtained from Eq. \ref{eq:huber_refit} are then used to modify the solar template and remaining sky background as shown below,

\begin{equation} \label{eq:solar_template_corr}
\begin{aligned}
        f_{\rm corr}(\lambda) &= f(\lambda - k_{\rm sky} + \Delta \lambda_{\rm median}) \\
        &+ m(\lambda - k_{\rm sky} + \Delta \lambda_{median}), \\
\end{aligned}
\end{equation}
where $f_{\rm corr}(\lambda)$ is the corrected solar template.

\begin{equation} \label{eq:skybg_corr}
\begin{aligned}
        b_{\rm corr}(\lambda) &= b + c(\lambda - k_{\rm sky} + \Delta \lambda_{\rm median}), \\
\end{aligned}
\end{equation}
where $b_{\rm corr}(\lambda)$ is the corrected sky background continuum.

The solar template, $f(\lambda - k_f)$, and remaining sky background continuum, $b$, in Eq. \ref{eq:skybg_model} are replaced by $f_{\rm corr}(\lambda)$ and $b_{\rm corr}(\lambda)$ respectively. The updated sky background model is then used in refitting to remove any residual, $r(\lambda)$. This refitting process can significantly reduce the median and standard deviation of the residual at each wavelength after a few iterations. Five iterations are used in this study.

\begin{figure}[ht]
\begin{center}
    \includegraphics[width=0.90\linewidth]{{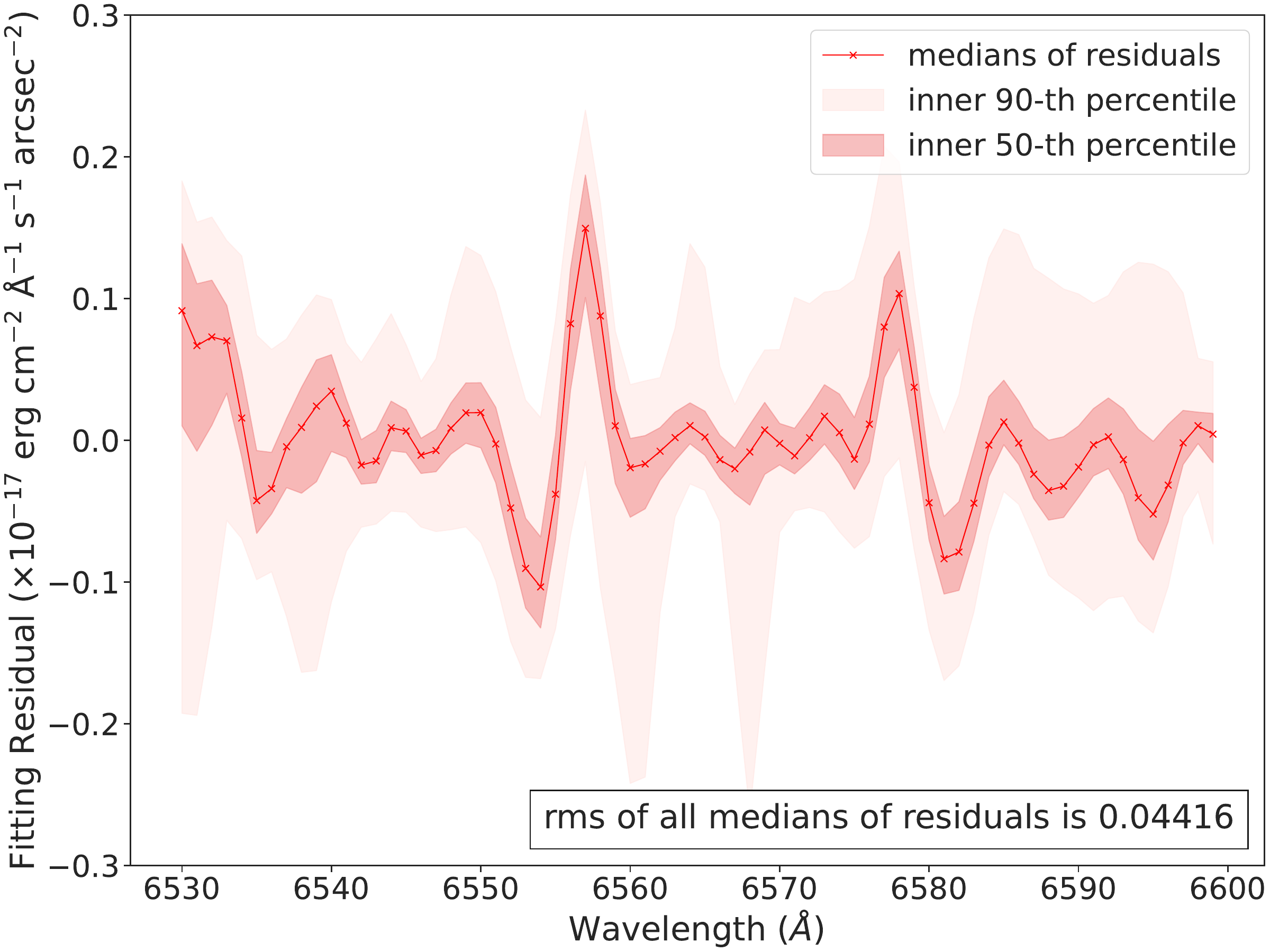}}
    \par\smallskip
    \includegraphics[width=0.90\linewidth]{{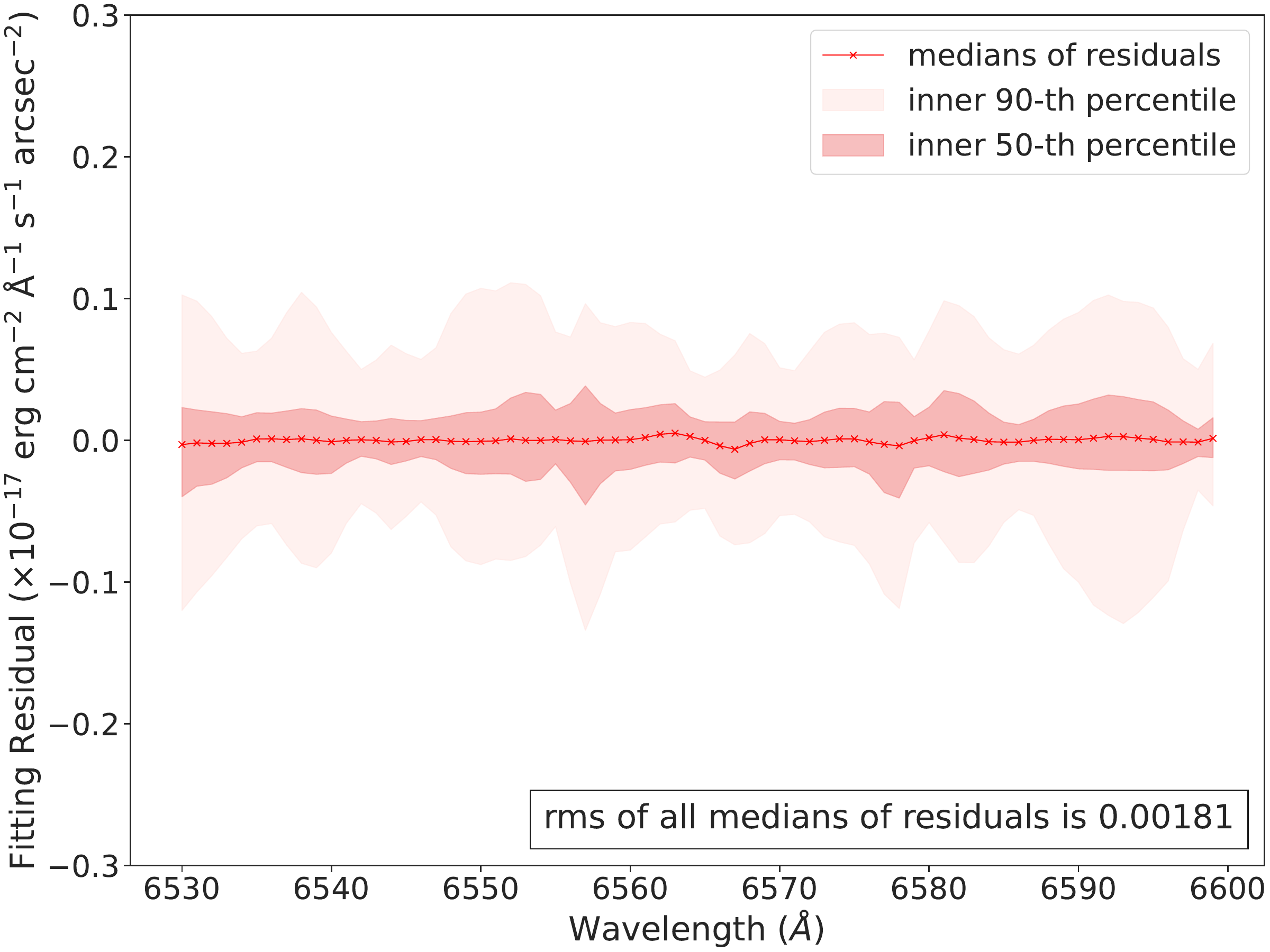}}
    \caption{Upper panel: Fitting residual of each exposure at each wavelength before the refitting process. Lower panel: Fitting residual of each exposure at each wavelength after the refitting process. The median of the fitting residual of all exposures at each wavelength (red crosses), the inner 50th percentile (dark red shading), and the inner 90th percentile (light red shading) are shown. The root mean square of all median residuals is given at the lower right of each panel.}
    \label{fig:refit_result}
\end{center}
\end{figure}

The refitting result is shown in Fig. \ref{fig:refit_result}. The median residuals are much closer to zero after refitting, and the root mean square of all medians of residuals at each wavelength decreases from 0.04416 to 0.00181 after refitting. The distribution of the residuals is more symmetric around the median as well.

\begin{figure}[h]
\begin{center}
    \includegraphics[width=0.90\linewidth]{{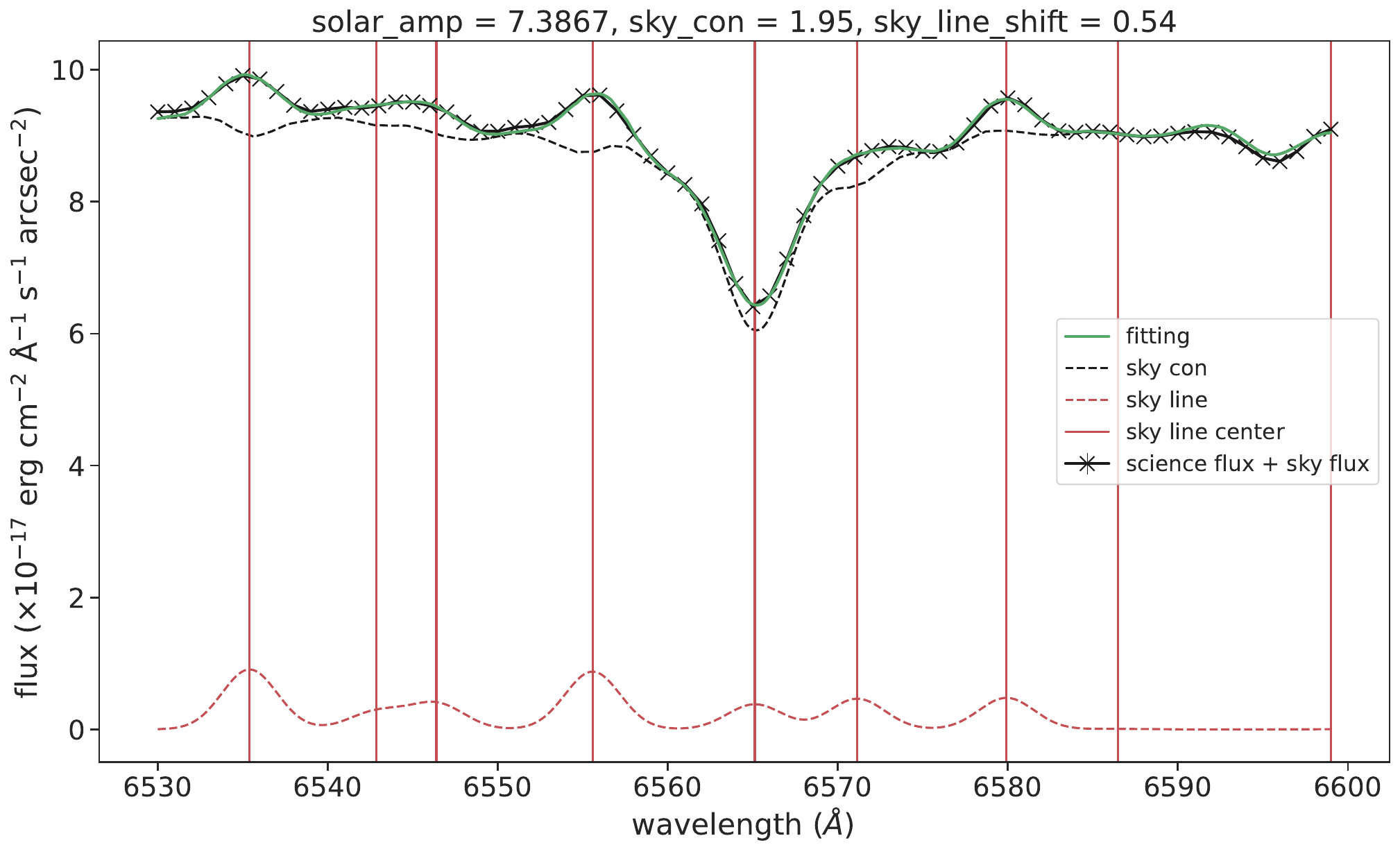}}
    \caption{Fitting result using the corrected solar template (plate 8051, modified Julian date 57057, MaStar id 27-1853, exposure id 193985). Fitted flux (green solid line), observed flux (black cross solid line with error bar), sky background continuum component (black dotted line), sky lines emission component (red dotted line) and line centres of sky lines (red vertical line) are plotted.}
    \label{fig:fitting_result_low}
\end{center}
\end{figure}

After this refitting process, we then have a final corrected sky background. An example of fitting result using the corrected sky background with Eq. \ref{eq:skybg_model} is shown in Fig. \ref{fig:fitting_result_low}.

\subsection{Geocoronal ${H\alpha}$ overview} \label{subsec:geoHa_overview}
Geocoronal \Ha\ emission originates from the excited hydrogen at the upper thermosphere and exosphere, which is mainly contributed by Lyman ${\rm \beta}$ excitation by solar radiation, and minor contribution from higher energy level (n > 3) cascade process that populates n = 3 (see Sect. 8 at \citet{mierkiewicz2006geocoronal} for details, also see \citet{Roesler2014geocoronal_halpha_atomic_mechanism}).

The geocoronal \Ha\ emission rate depends on many factors such as the strength of Lyman ${\rm \beta}$ flux, the profile of hydrogen density in the atmosphere, and viewing geometry \citep{Bishop1999lyeo_rt, Bishop2001geocoronal_Ha}. Especially, the geocoronal \Ha\ emission decreases with altitude as the density of hydrogen decreases \citep{Gardner_shadow_altitude}. The observed geocoronal \Ha\ intensity mostly captures the entire column of geocoronal \Ha\ emission along the line of sight.

\subsection{Shadow altitude overview} \label{subsec:shadow_alt_overview}
\begin{figure}[h]
\begin{center}
    \includegraphics[width=0.90\linewidth]{{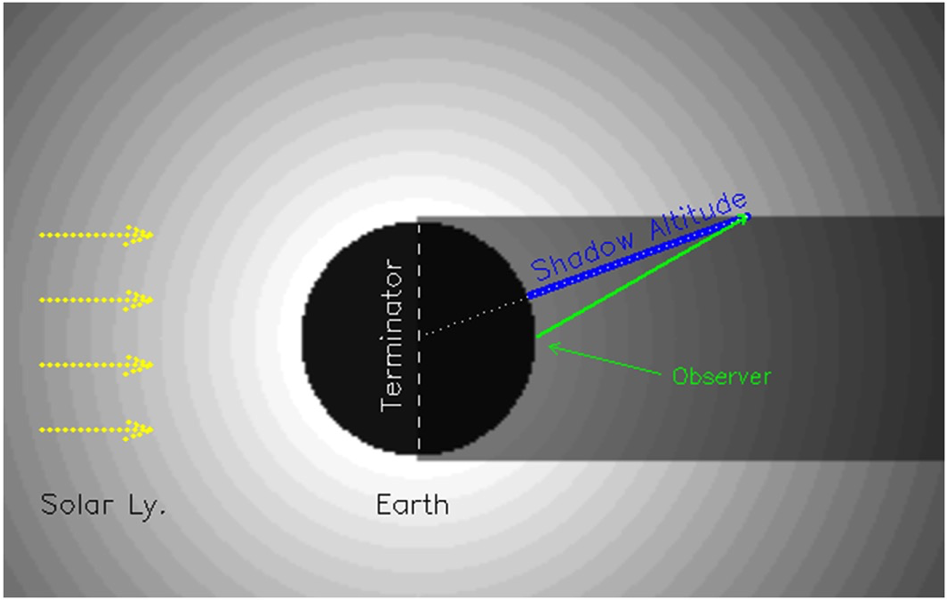}}
    \caption{Illustration of the observation geometry for an observer at the equator at local midnight, showing the shadow altitude (blue solid arrow), the line of sight (green arrow), and the shadow cone of the Earth (grey box; Fig. 1 of \citealt{Gardner_shadow_altitude}). This figure is an oversimplification because the plane formed by the observer's LOS and the centre of the Earth is not necessarily perpendicular to the terminator plane of the Earth. The actual geometry depends on the location of the observer, observation time, observation direction, and relative position of the Sun. It is generally a 3D calculation.}
    \label{fig:shadow_altitude}
\end{center}
\end{figure}

\begin{figure}[ht]
\begin{center}
    \includegraphics[width=0.90\linewidth]{{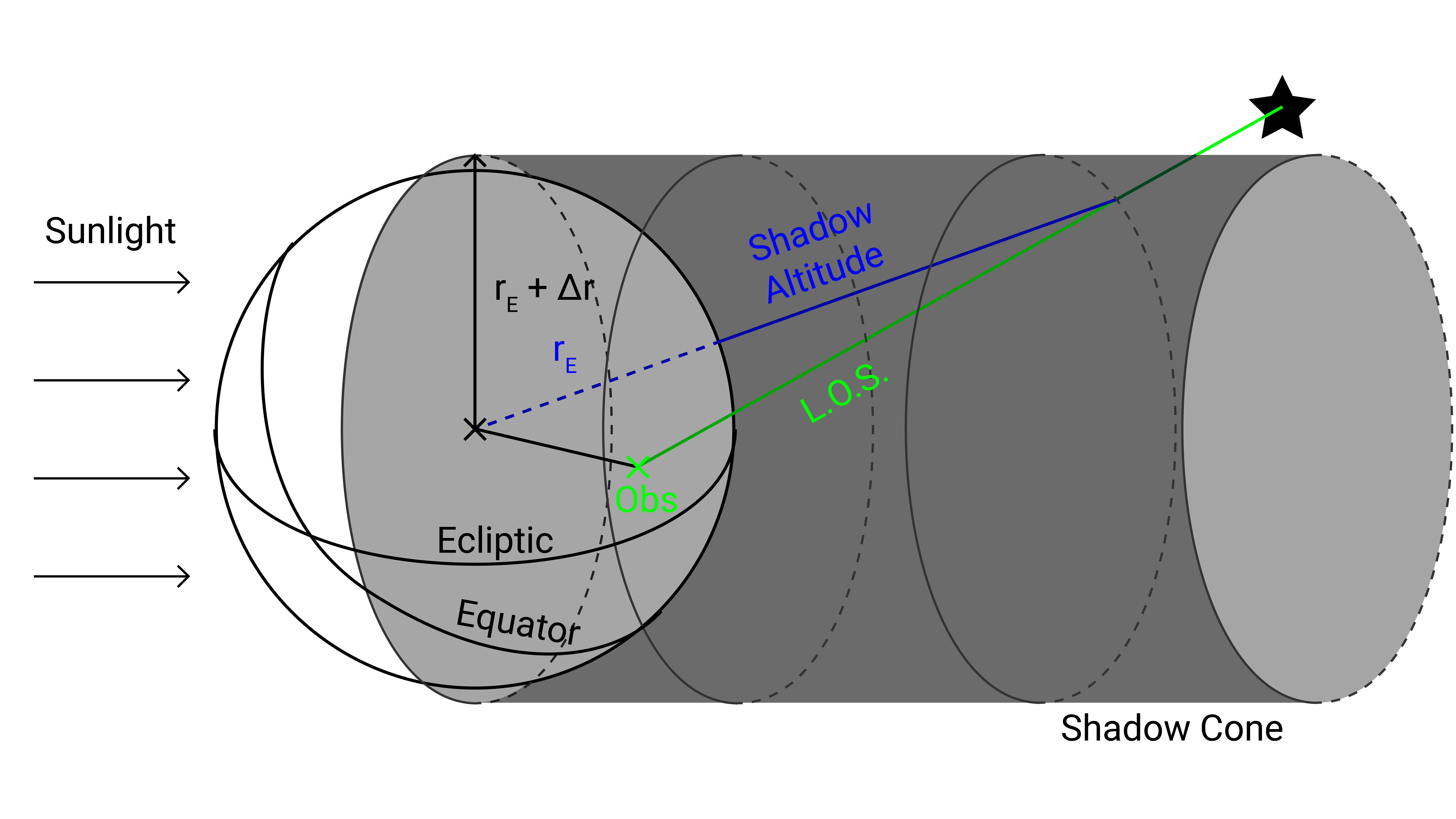}}
    \caption{3D illustration of the geometry in a similar manner of Fig. \ref{fig:shadow_altitude}. The shadow altitude (blue solid line), the radius of the Earth (blue dotted line), LOS (green dotted line), location of the observatory (green cross labelled as 'Obs'), and shadow cone (grey region) are shown. Notice the target (black star) is outside of the shadow cone, and the LOS is pointing into the paper in this illustration.}
    \label{fig:shadow_altitude_illustration}
\end{center}
\end{figure}

The shadow altitude is a geometric viewing parameter \citep{Gardner_shadow_altitude}, defined as the length from the surface of the Earth, in the radial direction, to the intersection between the line of sight (LOS) and the shadow cone of the Earth pointing, shown as the blue arrow in Fig. \ref{fig:shadow_altitude} and blue solid line in Fig. \ref{fig:shadow_altitude_illustration}. In general, the shadow altitude depends on the location of the observatory, observation time, and celestial coordinate (or equatorial coordinate, which can be easily converted by the observation time), as shown more clearly in Fig. \ref{fig:shadow_altitude_illustration}. We used a 3D vector calculation detailed in \citet{Guertault_2017} and implemented in \citet{Drory2023Shadow_alt_calculation} to calculate the shadow altitude. We additionally add 102 km to the base radius of the shadow cone since Lyman $\beta$ radiation would be fully absorbed by atmospheric ${\rm O_2}$ under 102 km \citep{Geocoronal_Halpha_Fabry-Perot_facility}.

The shadow altitude can be interpreted as the height of the base of the hydrogen emission column, and therefore related to the density of hydrogen at the top of the emission column, with a larger shadow altitude corresponding to a lower density of hydrogen. With this reasoning, we can characterise the observed geocoronal \Ha\ intensity with the shadow altitude of a particular line of sight.

\section{Results} \label{sec:results}

\subsection{Shadow altitude as a geocoronal \Ha\ flux estimator} \label{subsec:shadow_alt_fitting}

The geocoronal \Ha\ flux observed is defined to be the flux fitted from the sky background model in Eqs. \ref{eq:skybg_model} and \ref{eq:solar_template_corr}, ${\rm F(H\alpha)_{obs} \equiv \sqrt{2\pi}\sigma_{H\alpha}A_{H\alpha}}$. The uncertainty, ${\rm \delta F(H\alpha)_{obs}}$, is estimated as the diagonal term of the covariance matrix returned by \texttt{scipy.optimize.curve\_fit}.

We only use ${\rm F(H\alpha)_{obs}}$ of low Galactic \Ha\ pointings, assuming the detected \Ha\ is only coming from geocoronal sources. We further choose exposures with S/N of ${\rm F(H\alpha)_{obs}}$ > 3, which left us with 6305 exposures (86.4\,\% of the original exposures). Data points after the S/N cut are plotted against the shadow altitude, shown below in Fig. \ref{fig:geoHa_vs_shadow_altitude}.

\begin{figure}[ht!]
\begin{center}
    \includegraphics[width=0.90\linewidth]{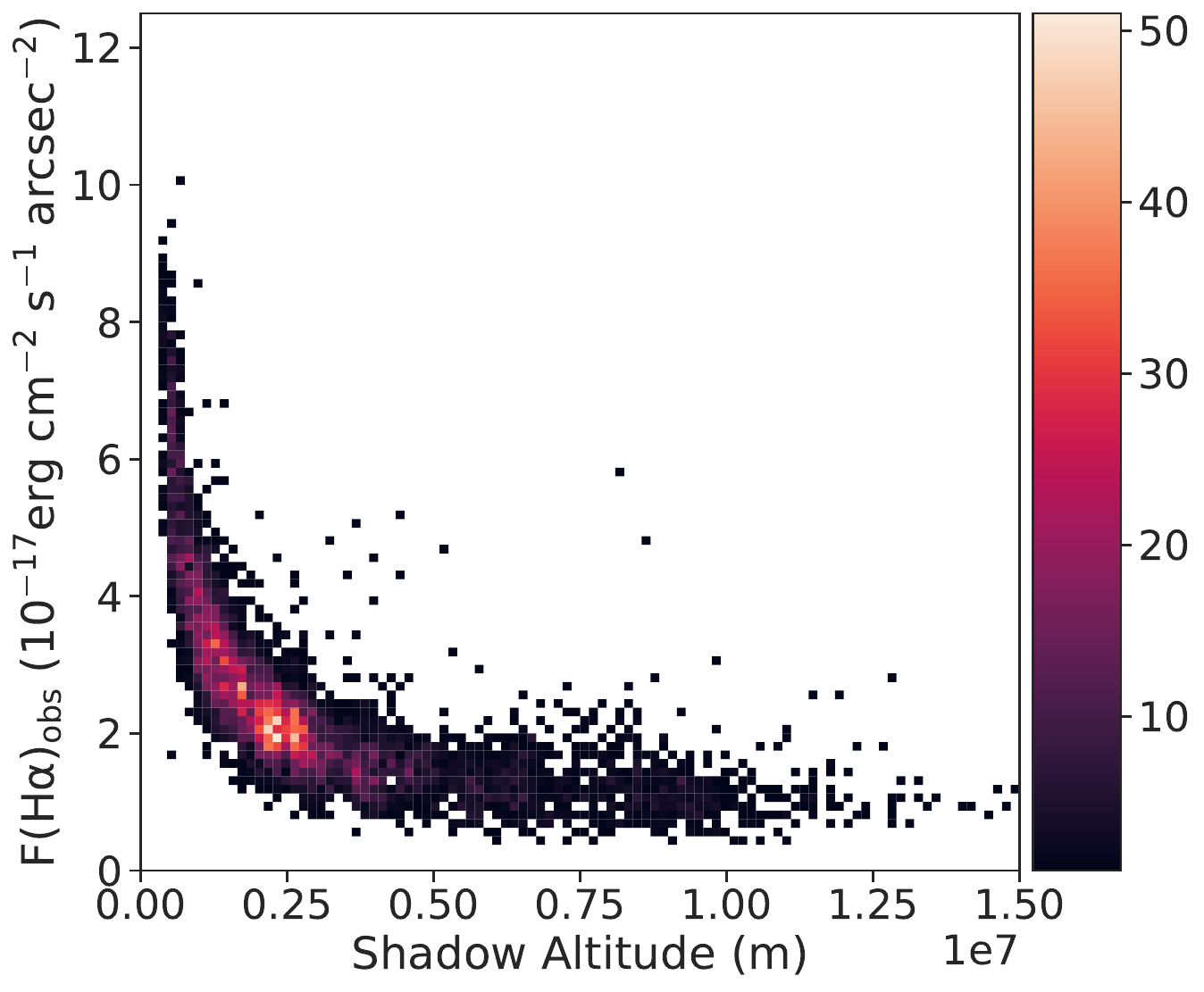}
    \caption{Distribution of ${\rm F(H\alpha)_{obs}}$ fitted from low Galactic \Ha\ IFUs after the S/N cut (y-axis) against the shadow altitude (x-axis). The colour bar represents the number density of data points.}
    \label{fig:geoHa_vs_shadow_altitude}
\end{center}
\end{figure}

A clear trend of decreasing geocoronal \Ha\ flux with increasing shadow altitude can be seen, this is expected as larger shadow altitude corresponds to a line of sight (LOS) nearly opposite from the Sun, with this LOS passing through the geocoronal layer at very high altitude and low gas density, and therefore less geocoronal \Ha\ emission. This result is consistent with the results found by WHAM \citep{Geocoronal_Halpha_Fabry-Perot_facility, mierkiewicz2006geocoronal, Gardner_shadow_altitude, https://doi.org/10.1029/2019JA026903}.

\begin{figure}[ht]
\begin{center}
    \includegraphics[width=0.90\linewidth]{{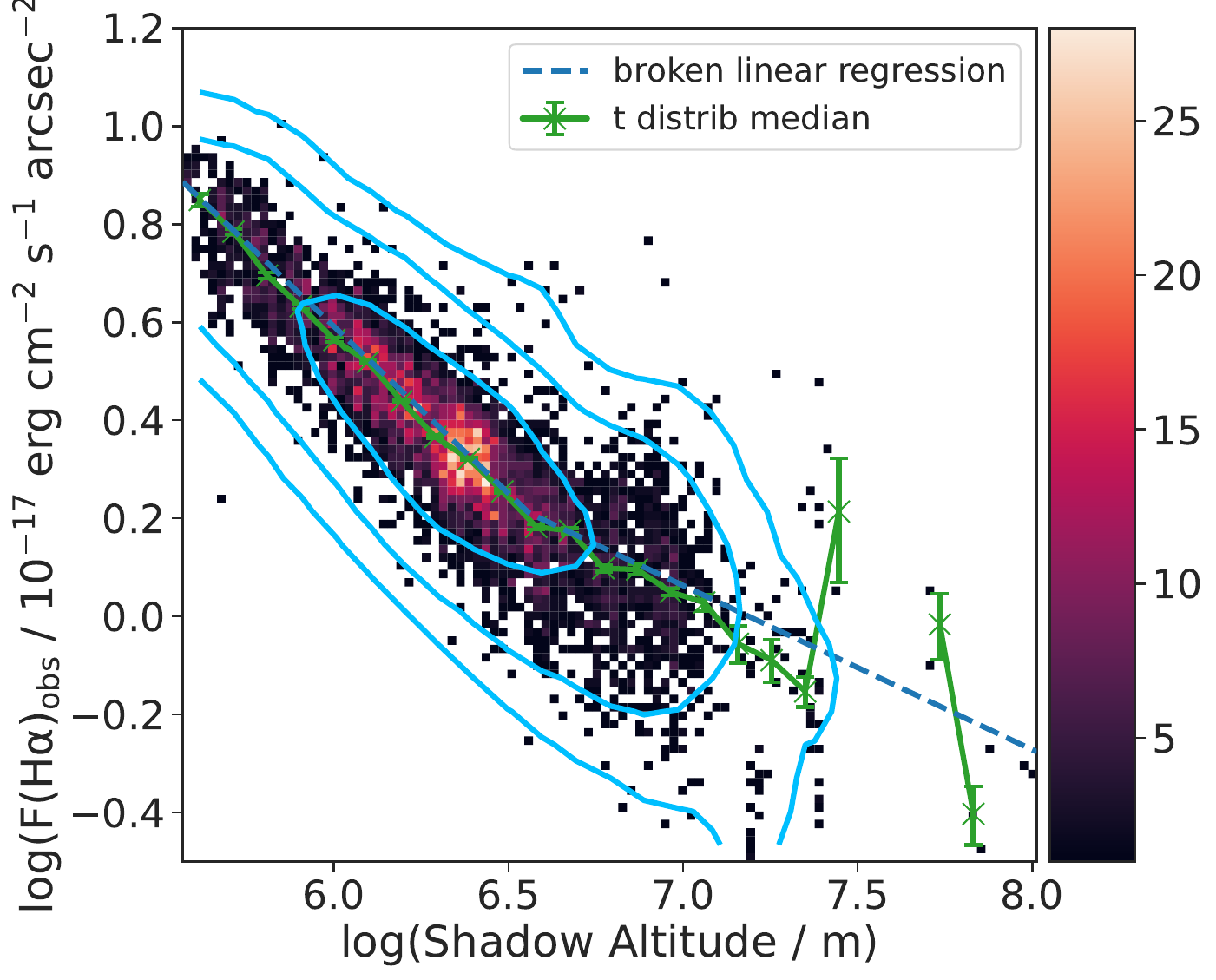}}
    \caption{The \Ha\ flux fitted from low Galactic \Ha\ pointing-exposures (geocoronal \Ha\ only) in log scale against the shadow altitude in log scale.
    Data points (pixels in the histogram) are fitted with a broken linear fitting (blue dashed line), binned median (green crosses with uncertainty of median estimated by t-distribution in linear space), and contour lines (light blue solid lines) at the 1 $\sigma$ (68.3\%), 2 $\sigma$ (95.5\%), and 3 $\sigma$ (99.7\%) levels are shown. The colour bar represents the number density of data points.}
    \label{fig:geoHa_vs_shadow_altitude_fitting}
\end{center}
\end{figure}

\begin{figure}[ht]
\begin{center}
    \includegraphics[width=0.90\linewidth]{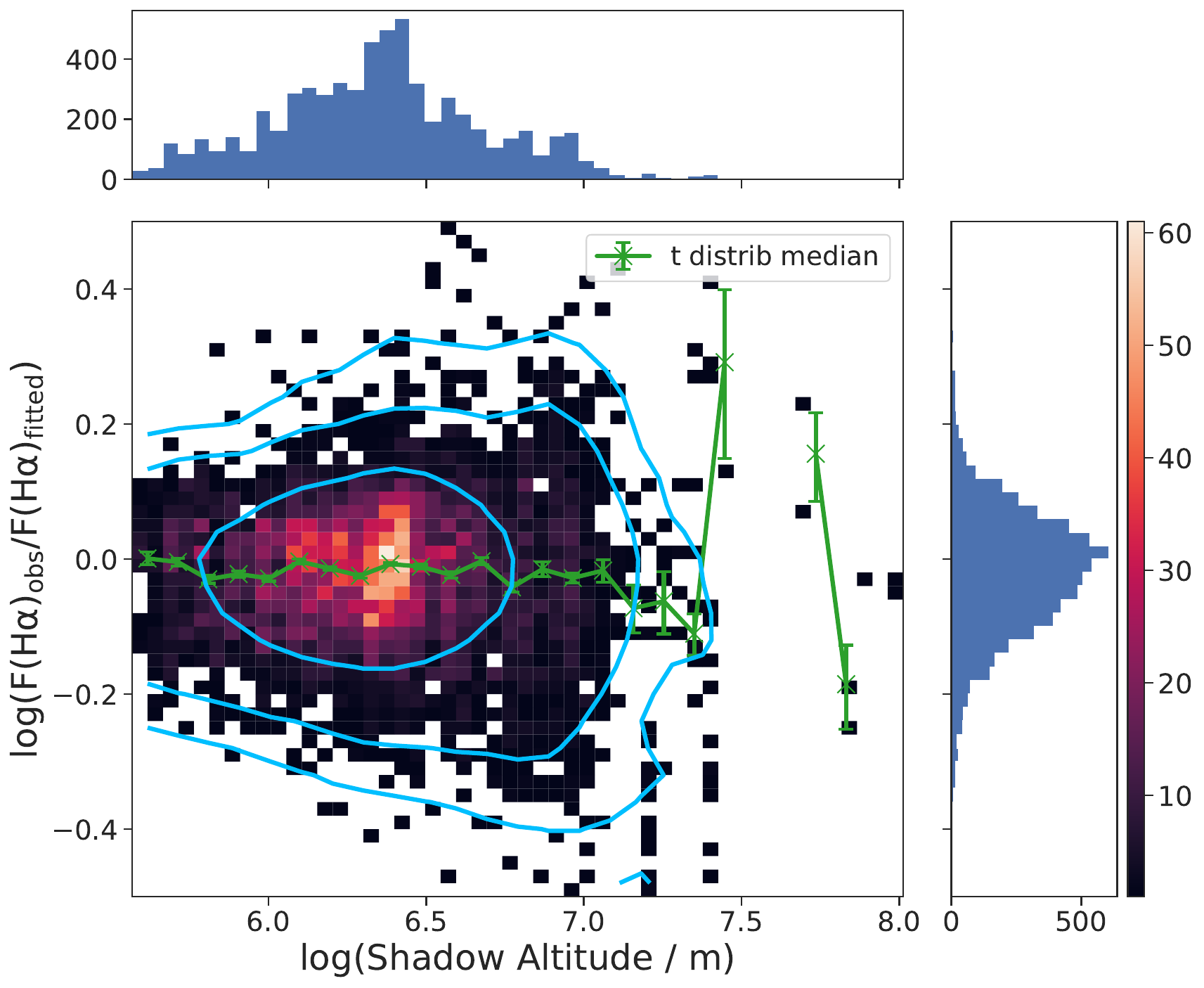}
    \caption{The residual (in log scale) of each IFU for the broken linear fit (pixels in the histogram) against shadow altitude. Binned medians (green crosses with uncertainty of median estimated by t-distribution in linear space) and contour lines (light blue solid lines) at the 1 $\sigma$ (68.3\%), 2 $\sigma$ (95.5\%), and 3 $\sigma$ (99.7\%) levels are shown. The colour bar represents the number density of data points.}
    \label{fig:residual_vs_shadow_altitude}
\end{center}
\end{figure}

We fit ${\rm F(H\alpha)_{obs}}$ as a function of shadow altitude using a two-segments broken linear model (\texttt{pwlf}, fitting package from \citealp{pwlf}) in the log-log scale, inverse variance weighting of ${\rm F(H\alpha)_{obs}}$ is used and the break point is treated as a free parameter. The fitting result is shown in Figs. \ref{fig:geoHa_vs_shadow_altitude_fitting} and \ref{fig:residual_vs_shadow_altitude} and summarised in Table \ref{table:shadow_alt_fit}. Physically, ${\rm F(H\alpha)_{obs}}$ never goes to zero because of multiple scattering of Lyman ${\rm \beta}$ in the atmosphere into Earth's shadow cone \citep{Gardner_shadow_altitude}, hence explaining the gentler slope at higher shadow altitude.

\begin{table}[htbp]
    \caption{Broken linear fitting of the shadow altitude method. Fitted coefficients for the logarithmic geocoronal H$\alpha$ flux in $10^{-17}\,\mathrm{erg\,cm^{-2}\,s^{-1}\,arcsec^{-2}}$ as a function of the logarithm of the shadow altitude in meters.}
    \label{table:shadow_alt_fit}
    \centering
    \small
    \begin{tabular}{l c c c}
        \hline\hline
        Segment & 1 & 2 &  \\
        \hline
        Slope & $-0.682$ & $-0.335$ & -- \\
        Intercept & $4.681$ & $2.408$ & -- \\
        Break/end point & $5.568$ & $6.558$ & $8.013$ \\
        \hline
    \end{tabular}
    \tablefoot{The slopes presented are in increasing shadow altitude direction.}
\end{table}

\subsection{Solar altitude method} \label{subsec:solar_alt_method}

The same problem of removing geocoronal \Ha\ at a medium spectral resolution was encountered in study based on the observations from LAMOST \citep{Zhang_2021}. They established a linear relationship between the flux ratio of geocoronal \Ha\ to OH$\lambda$6554 (6555.03\,\AA\ in vacuum) and the solar altitude (be mindful that the solar altitude is a measurement in terms of angles, whereas the shadow altitude is measured as a length).

We repeat their exercise with our data and use the same two segments broken linear fitting as before on high S/N of ${\rm F(H\alpha)_{obs}}$ IFUs detailed in Sect. \ref{subsec:shadow_alt_fitting}. An additional requirement of large VLSR was used in \citet{Zhang_2021}, which is not implemented here to provide a fair comparison between both methods.

\begin{figure}[ht!]
\begin{center}
    \includegraphics[width=0.90\linewidth]{{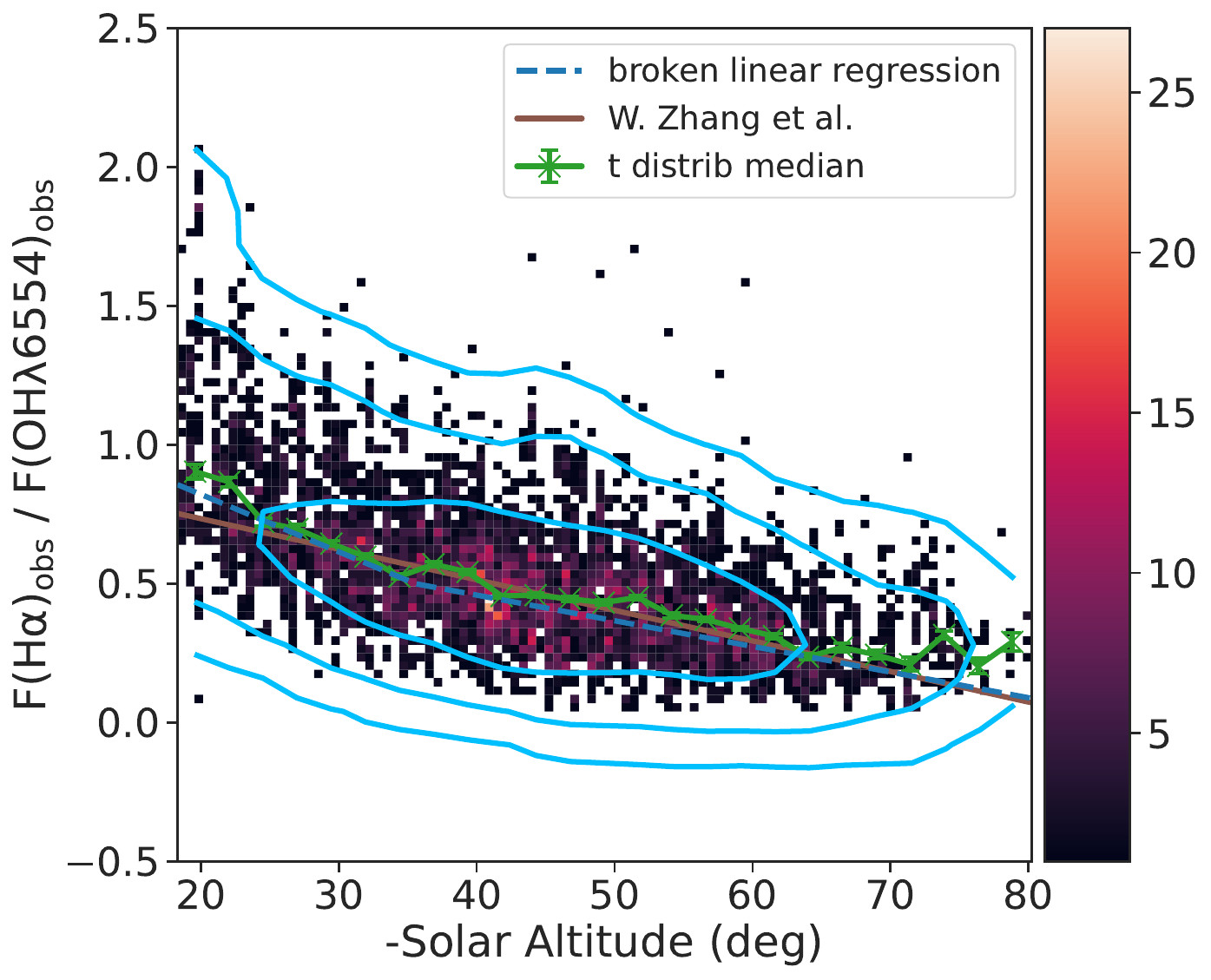}}
    \caption{The flux ratio of ${\rm F(H\alpha)_{obs}}$ to ${\rm F(OH\lambda6554)_{obs}}$ fitted from low Galactic \Ha\ pointings against the solar altitude. Fitting result from \citet{Zhang_2021} (brown solid line) is shown, other symbols are the same as those in Fig. \ref{fig:geoHa_vs_shadow_altitude_fitting}.}
    \label{fig:solar_altitude_fitting}
\end{center}
\end{figure}

\begin{figure}[ht!]
\begin{center}
    \includegraphics[width=0.90\linewidth]{{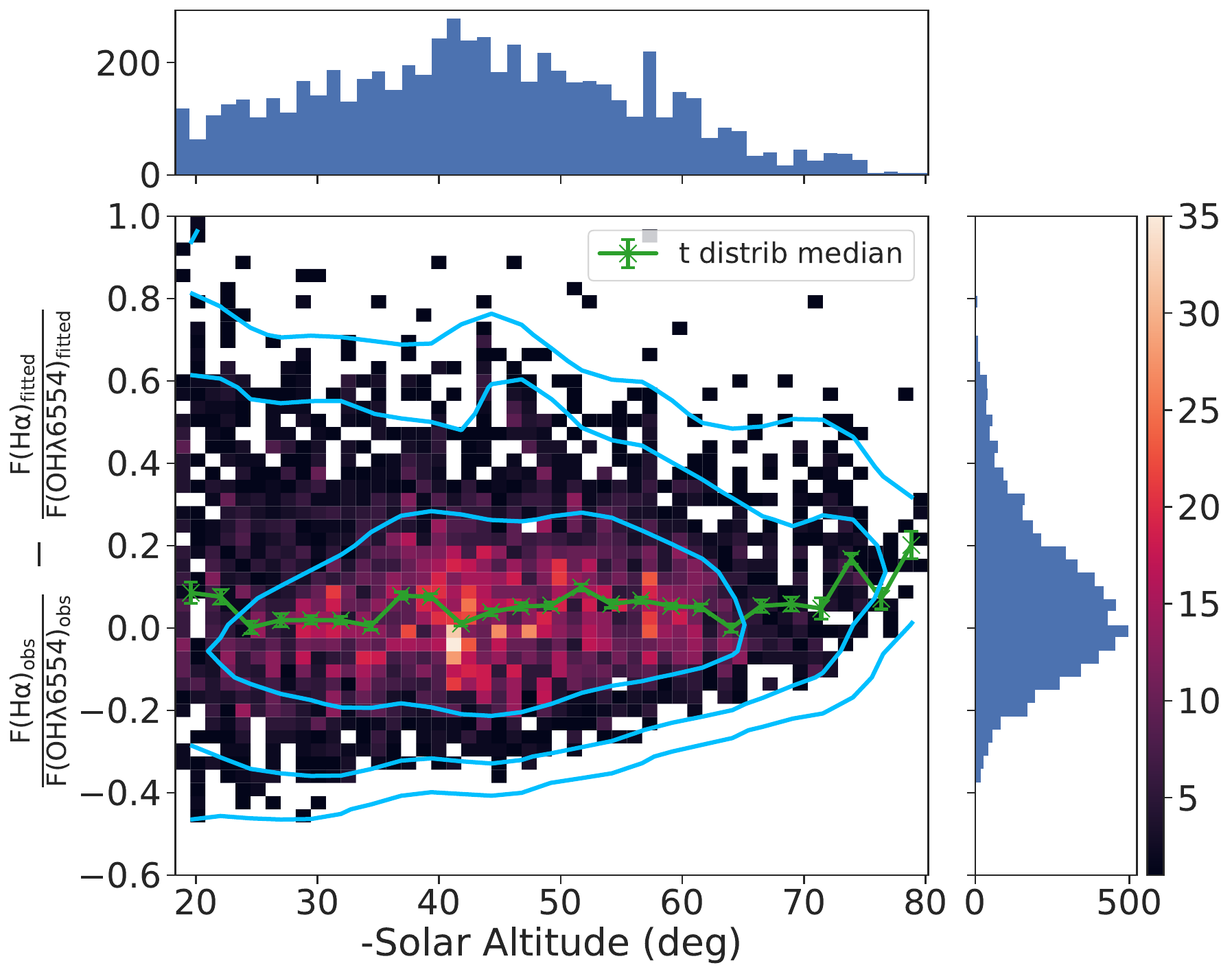}}
    \caption{The residual of broken linear fitting of the flux ratio ${\rm F(H\alpha)_{obs}}$ and ${\rm F(OH)_{obs}}$ fitted from low Galactic \Ha\ IFUs against the solar altitude. Symbols are the same as those in Fig. \ref{fig:residual_vs_shadow_altitude}.}
    \label{fig:residual_solar_altitude_fitting}
\end{center}
\end{figure}

The result of their method is tabulated in Table \ref{table:solar_alt_fit}, and plotted in Figs. \ref{fig:solar_altitude_fitting} and \ref{fig:residual_solar_altitude_fitting}. With our much larger data set, we find an overall consistent result with \citet{Zhang_2021}, except for small deviation at smaller solar altitudes.

\begin{table}[htbp]
    \caption{Fitted coefficients for the broken linear fitting of the flux ratio $\rm{F(\mathrm{H}\alpha)_\mathrm{obs}/F(\mathrm{OH}\lambda6554)_\mathrm{obs}}$ via the solar altitude method.}
    \label{table:solar_alt_fit}
    \centering
    \small
    \begin{tabular}{l c c c}
        \hline\hline
        Segment & 1 & 2 &  \\
        \hline
        Slope & $-0.0207$ & $-0.00924$ & -- \\
        Intercept & $1.235$ & $0.828$ & -- \\
        Break/end point & $18.300$ & $35.489$ & $80.228$ \\
        \hline
    \end{tabular}
    \tablefoot{The slopes presented are in decreasing solar altitude direction.}
\end{table}

\subsection{Results comparison} \label{subsec:comparison}

We compare the predictive power of two methods, $\delta$, which is defined using the root mean square of the fractional residual of all data points, shown below in Eq. \ref{eq:frac_uncer},

\begin{equation} \label{eq:frac_uncer}
\begin{aligned}
    \delta &= {\rm rms}\left(\frac{\Delta {\rm F(H\alpha)}} {\rm F(H\alpha)_{fitted}}\right),
\end{aligned}
\end{equation}
where $\Delta {\rm F(H\alpha) \equiv F(H\alpha)_{obs} - F(H\alpha)_{fitted}}$, and ${\rm F(H\alpha)_{fitted}}$ is the geocoronal \Ha\ flux predicted from the broken linear fitting of each method.

\begin{figure}[ht!]
\begin{center}
    \includegraphics[width=0.90\linewidth]{{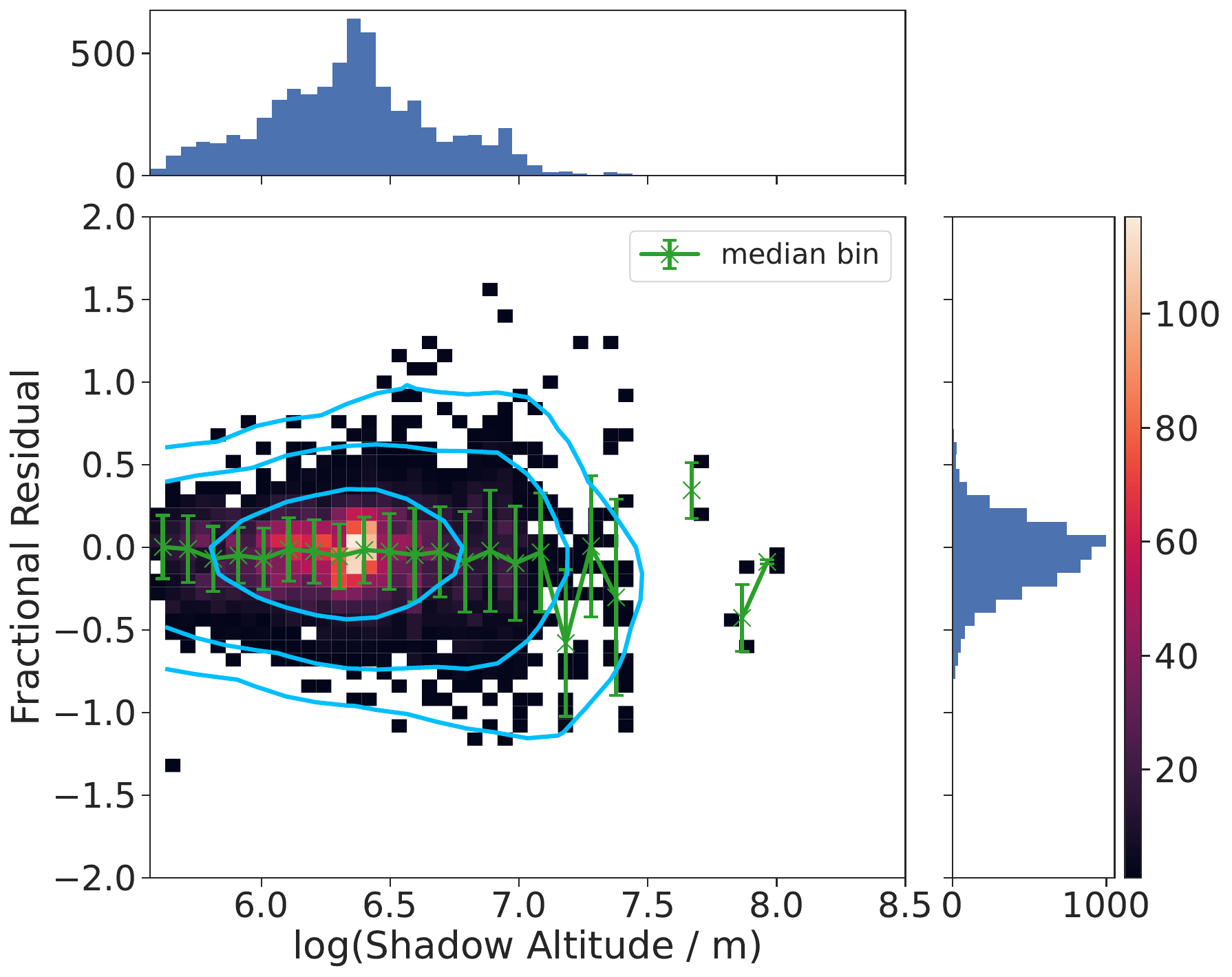}}
    \caption{The fractional residual from the prediction using shadow altitude from low Galactic \Ha\ IFUs. Symbols are the same as those in Fig. \ref{fig:residual_vs_shadow_altitude}.}
    \label{fig:frac_uncer_shadow_alt}
\end{center}
\end{figure}

\begin{figure}[ht!]
\begin{center}
    \includegraphics[width=0.90\linewidth]{{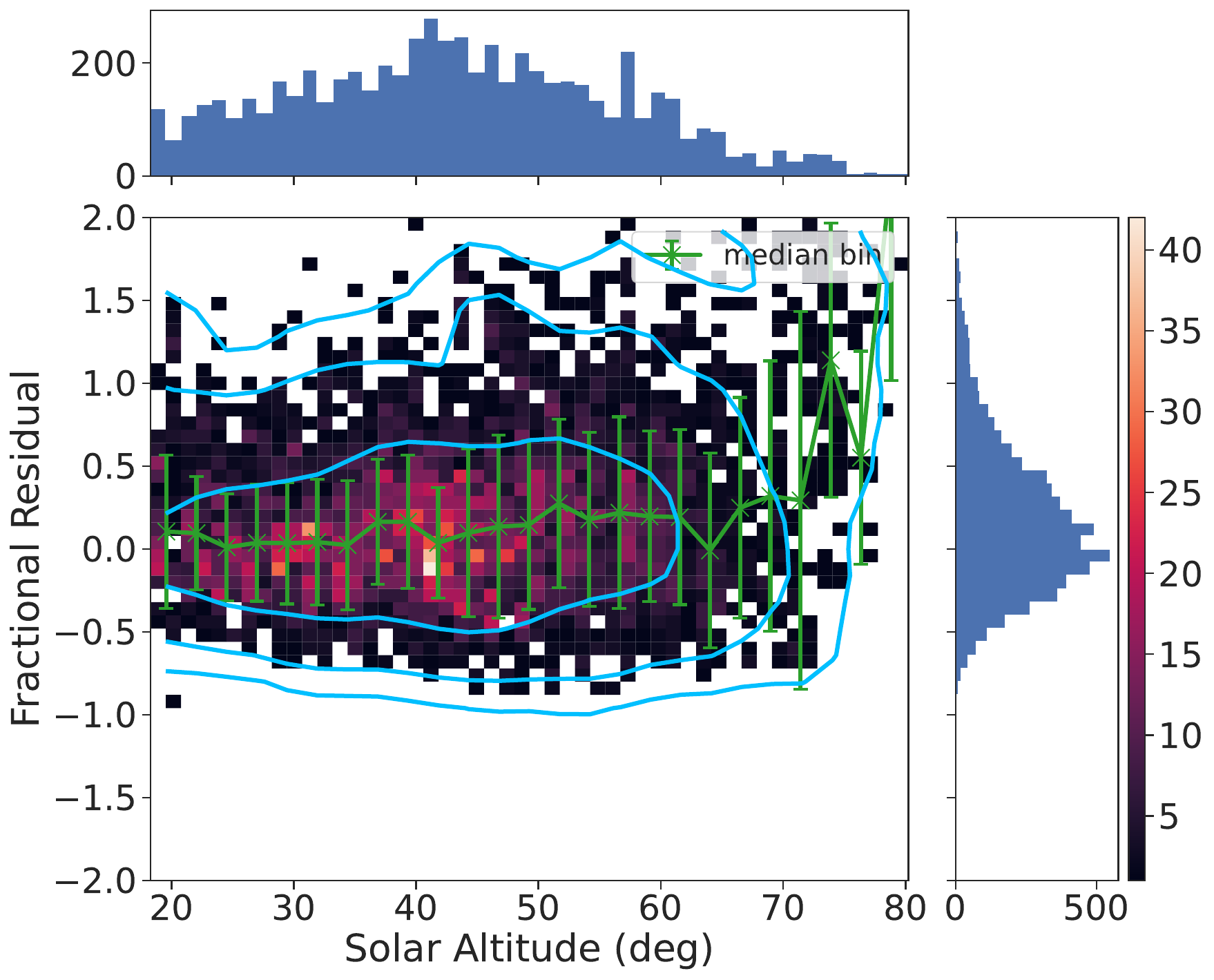}}
    \caption{The fractional residual from the prediction using \citet{Zhang_2021} method from low Galactic \Ha\ IFUs. Symbols are the same as those in Fig. \ref{fig:residual_vs_shadow_altitude}.}
    \label{fig:frac_uncer_solar_alt}
\end{center}
\end{figure}

The fractional residual of ${\rm F(H\alpha)_{obs}}$ of each method at different shadow altitudes or solar altitudes is also shown in Figs. \ref{fig:frac_uncer_shadow_alt} and \ref{fig:frac_uncer_solar_alt}. The fractional residuals of each method are fairly stable across different shadow altitudes or solar altitudes. Our method shows a more symmetric and smaller fractional residual distribution compared to the method of \citet{Zhang_2021}.

To determine if there is any intrinsic scatter for the models, we have also evaluated the root mean square of the ratio between the fitting residual, $\Delta {\rm F(H\alpha)}$, and the observed uncertainty of geocoronal \Ha\ flux, $\delta {\rm F(H\alpha)_{obs}}$, shown below in Eq. \ref{eq:intrinsic scatter},

\begin{equation} \label{eq:intrinsic scatter}
\begin{aligned}
    {\rm rms}
    \left(\frac {\Delta {\rm F(H\alpha)}}{\rm \delta F(H\alpha)_{obs}}\right).
\end{aligned}
\end{equation}
This number should be unity if the fitting residual, $\Delta {\rm F(H\alpha)}$, is limited by the uncertainty of the data points, $\delta {\rm F(H\alpha)_{obs}}$.

The result is tabulated in Table \ref{table:frac_uncer_result}. The error analysis is detailed in the appendix \ref{sec:delta_motivation}.

\begin{table}[htbp]
    \caption{Comparison of fractional uncertainty results between the shadow altitude method and the solar altitude method from \citet{Zhang_2021}.}
    \label{table:frac_uncer_result}
    \centering
    \small
    \begin{tabular}{l c c}
        \hline\hline
        Parameter & Shadow altitude & \cite{Zhang_2021} \\
        \hline
        $\delta$ & $24.41\%$ & $55.27\%$ \\
        \noalign{\smallskip}
        $\mathrm{rms}\left( \frac{\Delta F(\mathrm{H}\alpha)}{\delta F(\mathrm{H}\alpha)_\mathrm{obs}} \right)$ & $3.061$ & $5.373$ \\
        \hline
    \end{tabular}
    \tablefoot{
        $\delta$ represents the root mean square of the fractional residual. The second row shows the standard deviation of the ratio between the residual $\Delta F$ and the observational error $\delta F$.
    }
\end{table}

The root mean square of the fractional residual, $\delta$, from our method is about half of that based on the method of \citet{Zhang_2021}. Moreover, both methods show a $\mathrm{rms}(\Delta \mathrm{F(H\alpha)} / \delta \mathrm{F(H\alpha)_{obs}}) > 1$, indicating that intrinsic scatter is still present and contributing to $\delta$.

\section{Discussion} \label{sec:discussion}
\subsection{Potential sources of intrinsic scatter} \label{subsec:intrinsic_scatter_sources}
Despite the success of predicting the geocoronal \Ha\ flux from shadow altitude, the intrinsic scatter is still a major contributing factor to our standard deviation of fractional residual, $\delta$. We discuss some of the potential contributors to the intrinsic scatter.

\subsubsection{Line-of-sight geometry}
\begin{figure}
\begin{center}
    \includegraphics[width=0.90\linewidth]{{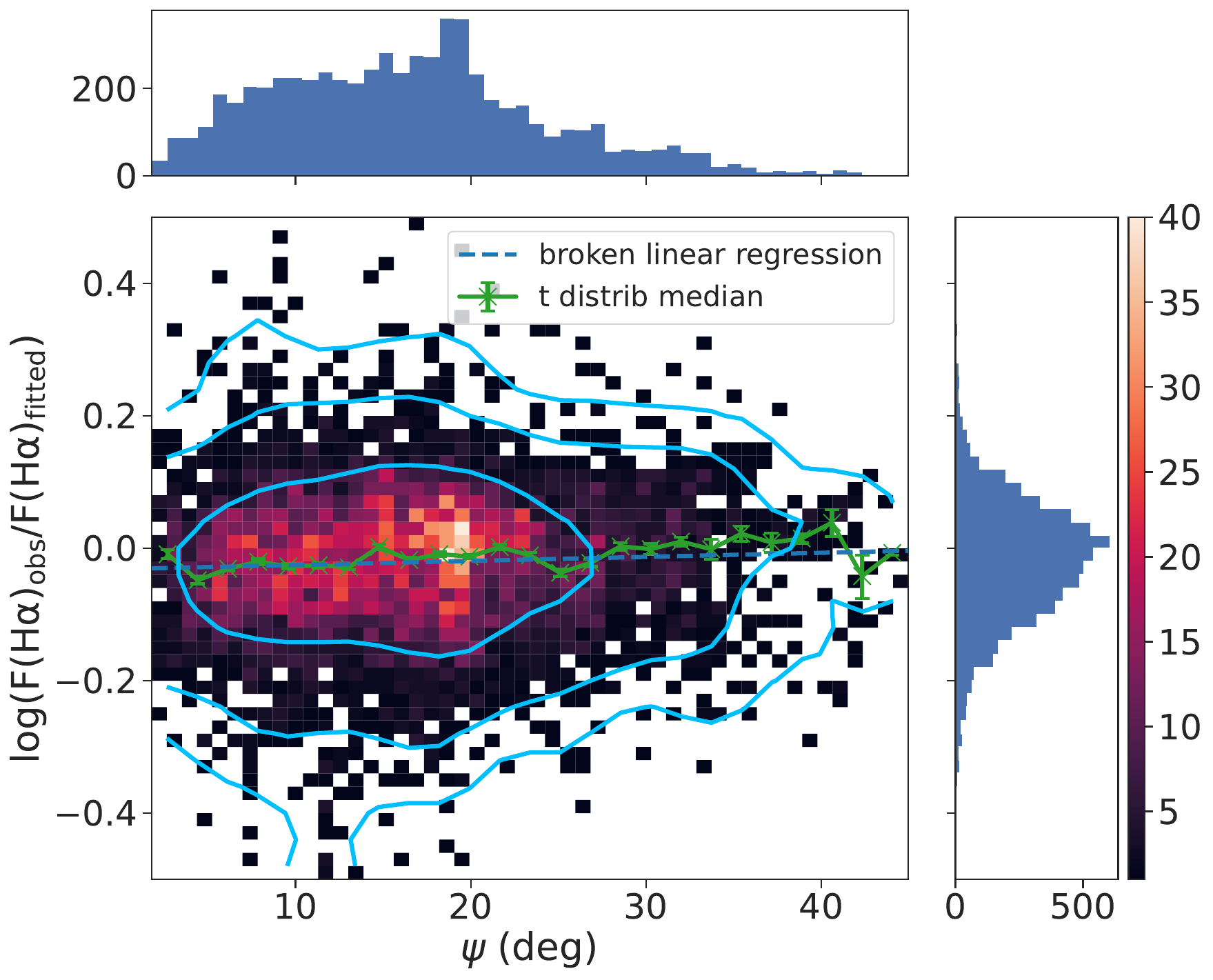}}
    \caption{The residual of the broken linear fit of Fig. \ref{fig:geoHa_vs_shadow_altitude_fitting} of each IFU against ${\rm \psi}$, the angle between the shadow altitude line and the LOS. Symbols are the same as those in Fig. \ref{fig:geoHa_vs_shadow_altitude_fitting}.}
    \label{fig:residual_vs_psi}
\end{center}
\end{figure}

One may naturally question whether, since different LOS correspond to different emission columns sharing the same shadow altitude, the shadow altitude is the only geometric parameter affecting the observed intensity of geocoronal \Ha. This is also discussed in \cite{Geocoronal_Halpha_Fabry-Perot_facility}, which suggested a variation of $\sim$ 1 Rayleigh, and up to 2 Rayleigh at low shadow altitude. We investigate this by plotting the residual of the broken linear fitting against the angle between the shadow altitude line (blue solid line in Fig. \ref{fig:shadow_altitude_illustration}) and LOS at the shadow cone surface, ${\rm \psi}$, in Fig. \ref{fig:residual_vs_psi}. However, no significant relation between the residual and ${\rm \psi}$ can be found. This is somewhat expected, given that the emission column will only change significantly at extremely large ${\rm \psi}$, which is rare in our data.

\subsubsection{Solar activity}
Solar activity could contribute to the intrinsic scatter. The amount of UV flux from the Sun will affect the strength of geocoronal \Ha, ${\rm F(H\alpha)_{obs}}$, as it originates from Lyman $\beta$ excitation by sunlight. We investigated the effect of the 11-year solar cycle, the Lyman $\beta$ flux from the solar chromosphere, and the total solar irradiance (TSI) on ${\rm F(H\alpha)_{obs}}$.

\begin{figure}[ht!]
\begin{center}
    \includegraphics[width=0.90\linewidth]{{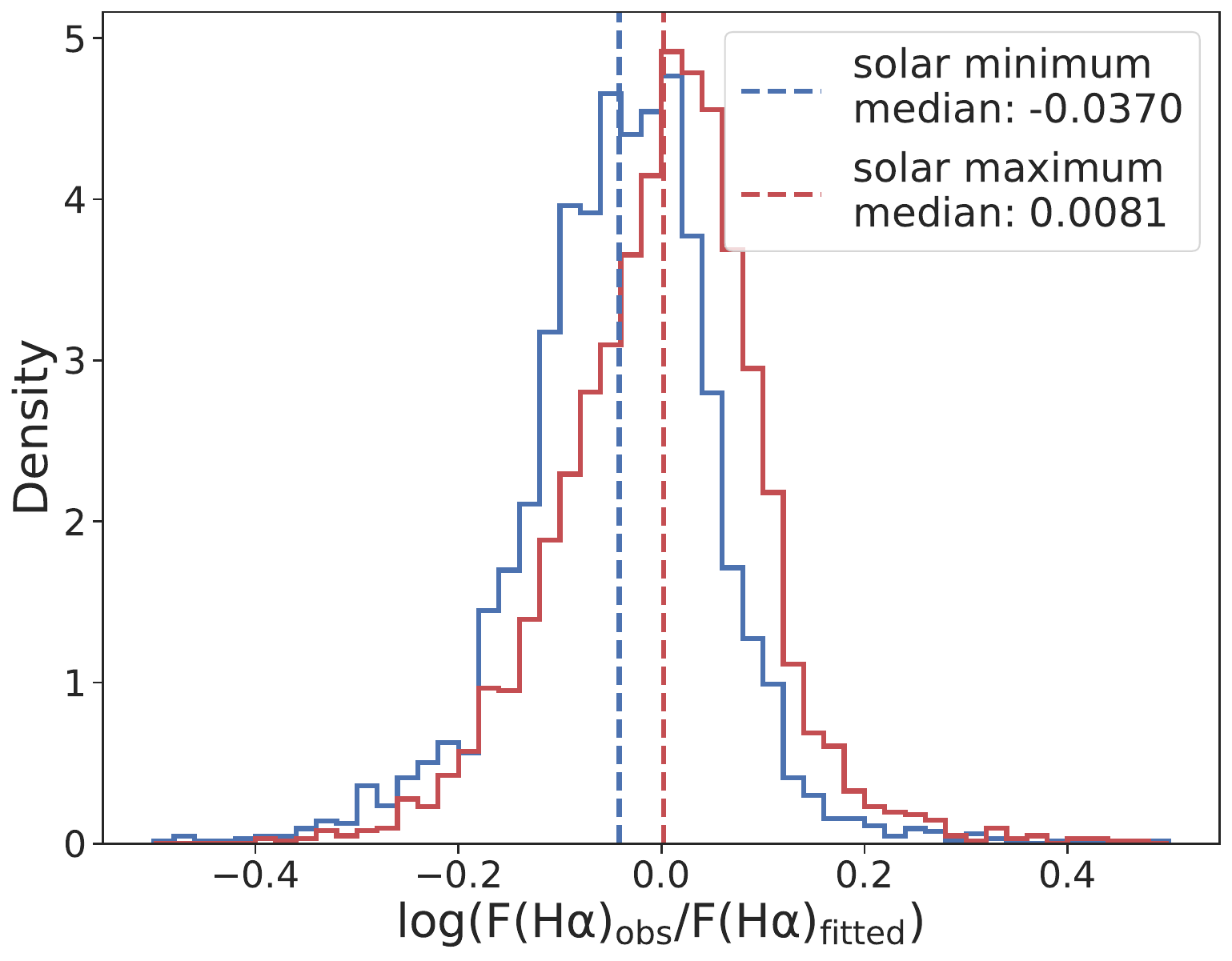}}
    \par\smallskip
    \includegraphics[width=0.90\linewidth]{{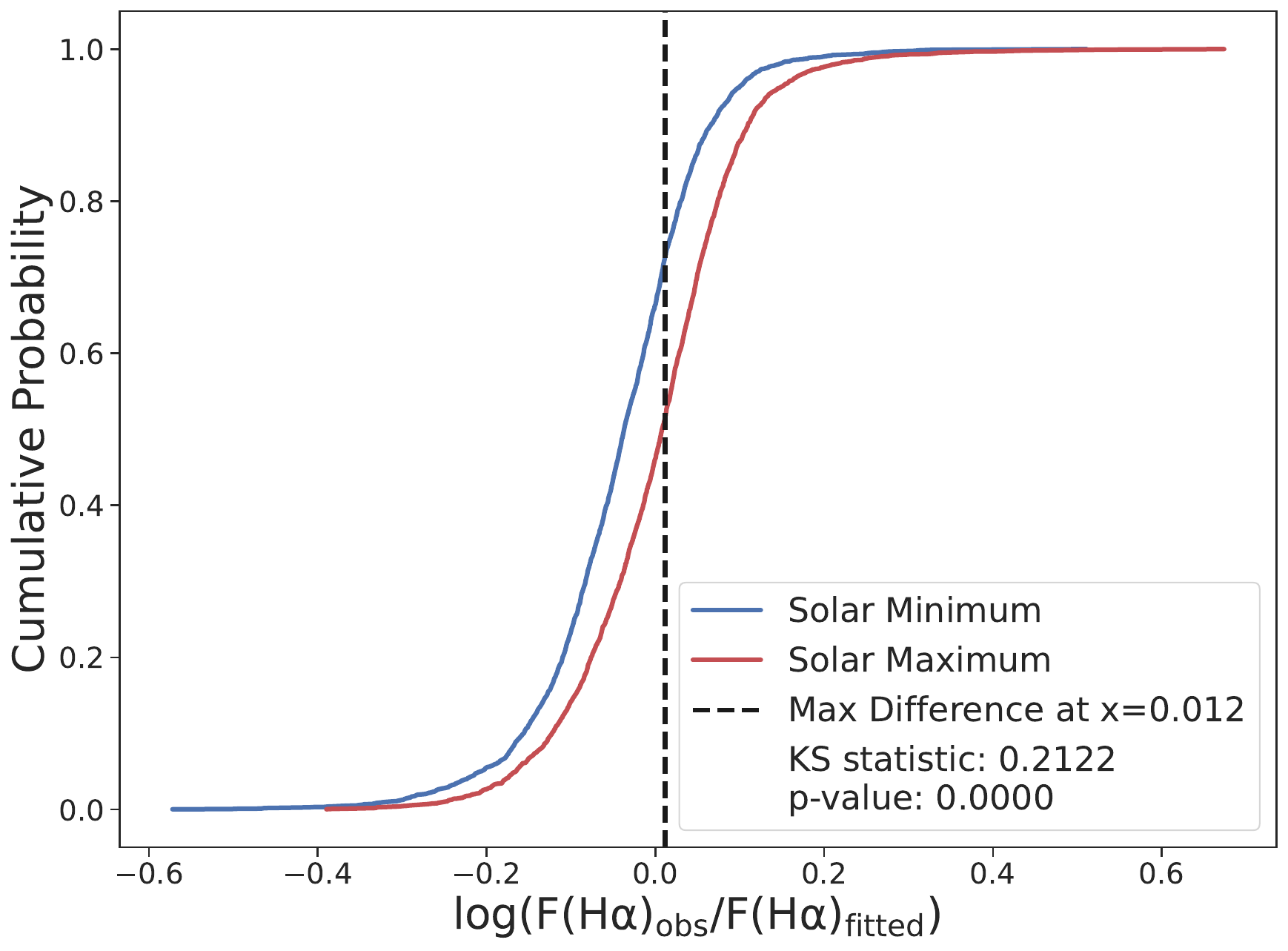}}
    \caption{Upper panel: Normalised distribution of the fitting residual in log space. Lower panel: Cumulative distribution of the fitting residual in log space. The blue and red solid lines represent data obtained after June 2017 (solar minimum) and before June 2017 (solar maximum), respectively. In the upper panel, the dashed lines represent the median fitting residuals. In the lower panel, the black dashed line marks the maximum separation between the two datasets.}
    \label{fig:geoHa_solar_cycle}
\end{center}
\end{figure}

The MaStar data we use span from 11 December 2014 to 30 July 2020, which roughly corresponds to the second half of the solar cycle 24. We split the data set into before and after 1 June 2017, corresponding to the solar maximum period and solar minimum period respectively. We find that there is indeed a stronger geocoronal \Ha\ flux during the solar maximum period compared to that of the solar minimum as expected. A simple Kolmogorov-Smirnov (K-S) test shows that the probability that the two subsamples come from the same parent distribution is almost zero. The result is shown in Fig. \ref{fig:geoHa_solar_cycle}. It is also noted that MaStar data only spans the second half of a single, relatively weak solar cycle. We expect the impact of solar activity on the residual, $\Delta {\rm F(H\alpha)}$, to be larger during a stronger solar cycle. However, this remains untested for now.

\begin{figure}[ht!]
    \begin{center}
        \includegraphics[width=0.90\linewidth]{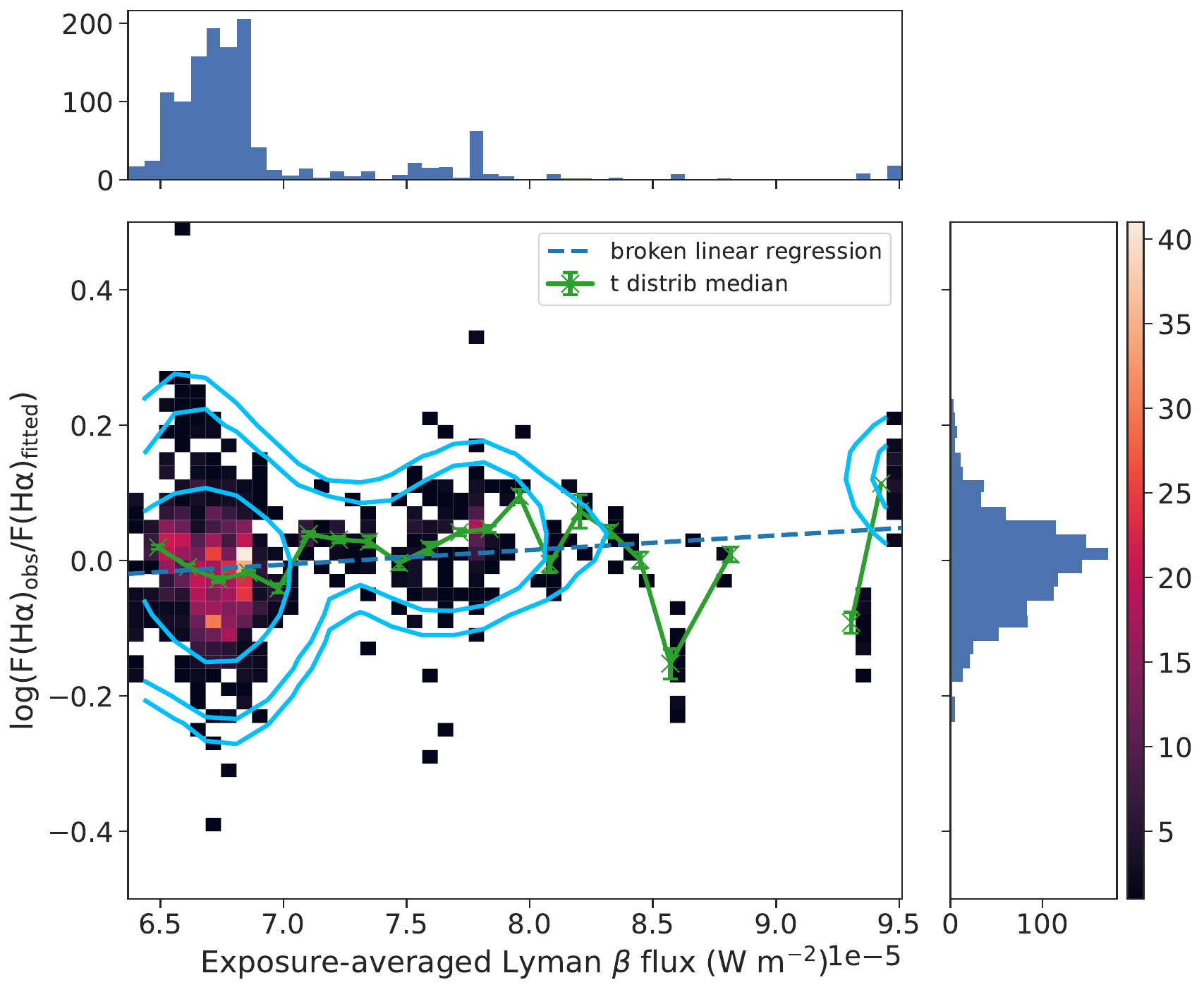}
    \caption{The residual of the broken linear fit of Fig. \ref{fig:geoHa_vs_shadow_altitude_fitting} of each IFU against exposure averaged Lyman $\beta$ flux. Symbols are the same as those in Fig. \ref{fig:geoHa_vs_shadow_altitude_fitting}.}
    \label{fig:residual_vs_lyman_beta}
    \end{center}
\end{figure}

For the Lyman $\beta$ flux, we use the 1 minute cadence data obtained by NASA Solar Dynamics Observatory Extreme Ultraviolet Variability Experiment \citep{Woods2012NASA_SDO}, with the level 4 lines fit data files including the Gaussian line flux provided by the laboratory for atmospheric and space physics in Boulder, Colorado \citep{woods2026revealingflareenergeticsdynamics}. For each MaStar exposure, we measure the average Lyman $\beta$ flux within the exposure window. There are 1304 exposures with at least one Lyman $\beta$ flux measurement during its exposure. The effect of the average Lyman $\beta$ flux on the fitted residual of ${\rm F(H\alpha)_{obs}}$ is shown in Fig. \ref{fig:residual_vs_lyman_beta}. Indeed the residual of ${\rm F(H\alpha)_{obs}}$ is larger at higher Lyman $\beta$ flux. The trend is weak when compared to the dispersion of the residual ($R^2$ is 0.0234). When using a two-segment multivariate broken linear fitting of ${\rm log_{10}(F(H\alpha)_{obs})}$ with both ${\rm log_{10}(shadow\ altitude)}$ and Lyman $\beta$ flux as predictors, $\delta$ decreases meaningfully to 18.02\%. However, given that continuous Lyman $\beta$ observation is only available for a short time window per day, and only a small fraction of MaStar exposures coincide with the Lyman $\beta$ observation window, we do not use Lyman $\beta$ flux predictor in our base model.

\begin{figure}[ht!]
    \begin{center}
        \includegraphics[width=0.90\linewidth]{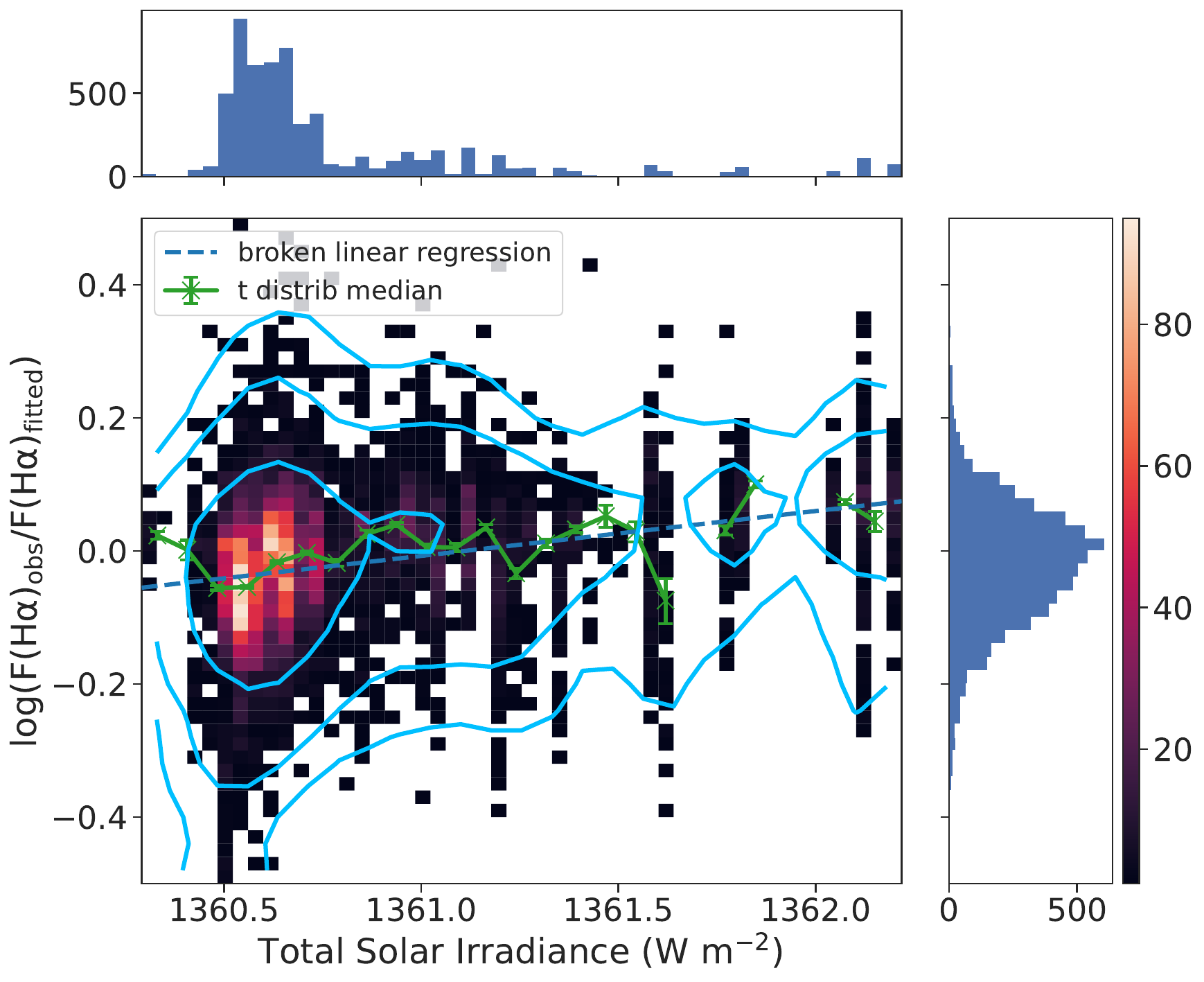}
    \caption{The residual of the broken linear fit of Fig. \ref{fig:geoHa_vs_shadow_altitude_fitting} of each IFU against total solar irradiance (TSI). Symbols are the same as those in Fig. \ref{fig:geoHa_vs_shadow_altitude_fitting}.}
    \label{fig:residual_vs_TSI}
    \end{center}
\end{figure}

For the TSI, we use the hourly data obtained by the Variability of Solar Irradiance and Gravity Oscillations experiment on the cooperative ESA/NASA Mission SOlar and Heliospheric Observatory (version 8.0, using PMO6V-A+B(Fused)) with the new scale \citep{finsterle2021total}. The effect of TSI on the fitted residual of ${\rm F(H\alpha)_{obs}}$ with shadow altitude is shown in Fig. \ref{fig:residual_vs_TSI}. We find an overall increasing trend of the residual of shadow altitude fitting with the TSI. However, the trend is weak when compared to the dispersion of the residual (slope obtained is 0.0674, and $R^2$ is 0.0612). A two-segment multivariate broken linear fitting of ${\rm log_{10}(F(H\alpha)_{obs})}$ with both ${\rm log_{10}(shadow\ altitude)}$ and TSI together is also tested, $\delta$ decreases slightly from 24.41\% to 23.42\%. Hence, we conclude that the TSI alone is not enough to explain all the intrinsic scatter of ${\rm F(H\alpha)_{obs}}$ observed.

\subsubsection{Earth--Sun distance}
The distance between the Sun and the Earth varies by $\sim$ 3.3\% within a year, with perihelion in early January and aphelion in early July. This causes both the solar radiation received by the Earth and the observed geocoronal \Ha\ flux, ${\rm F(H\alpha)_{obs}}$, to vary.

\begin{figure}[ht]
\begin{center}
    \includegraphics[width=0.90\linewidth]{{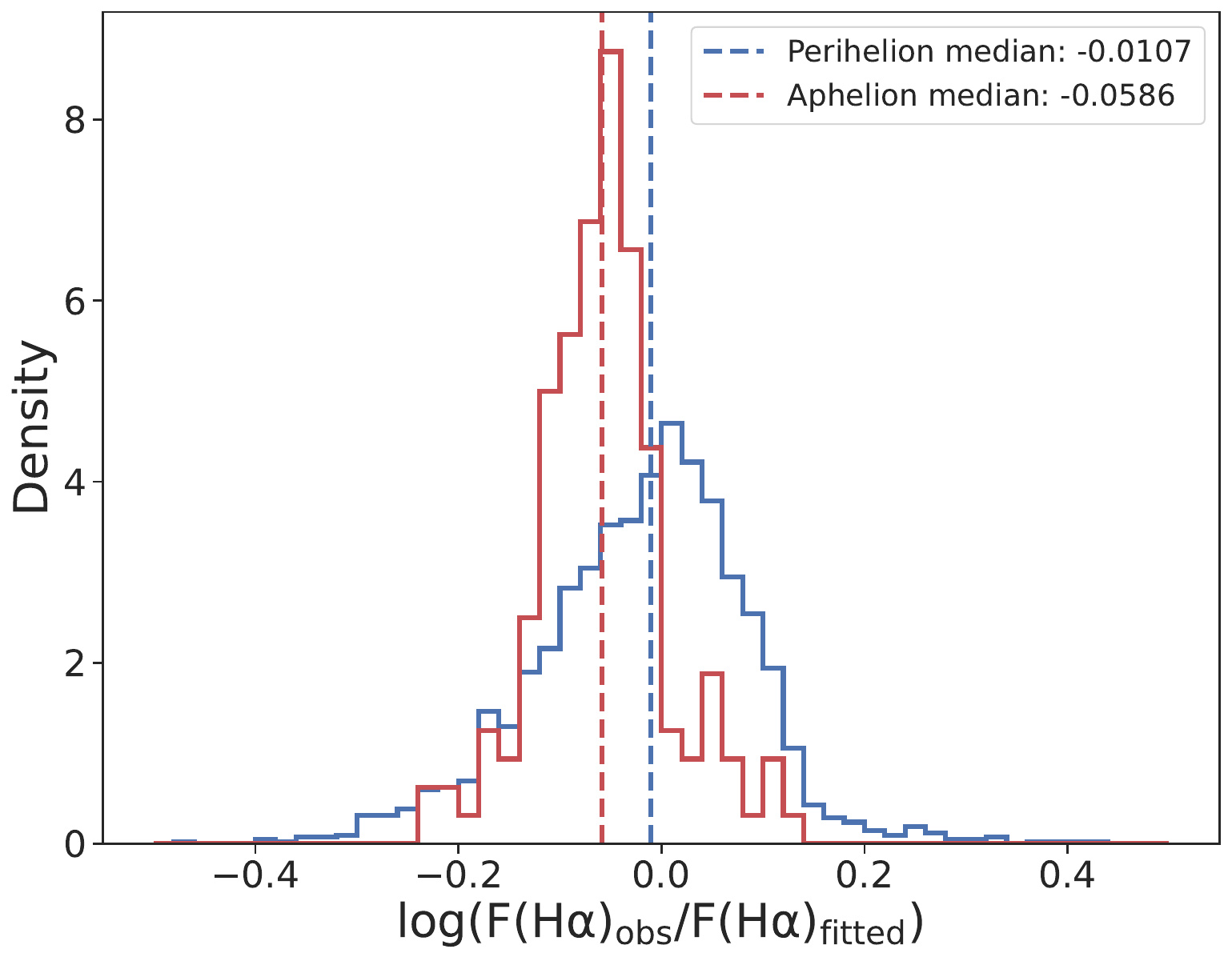}}
    \par\smallskip
    \includegraphics[width=0.90\linewidth]{{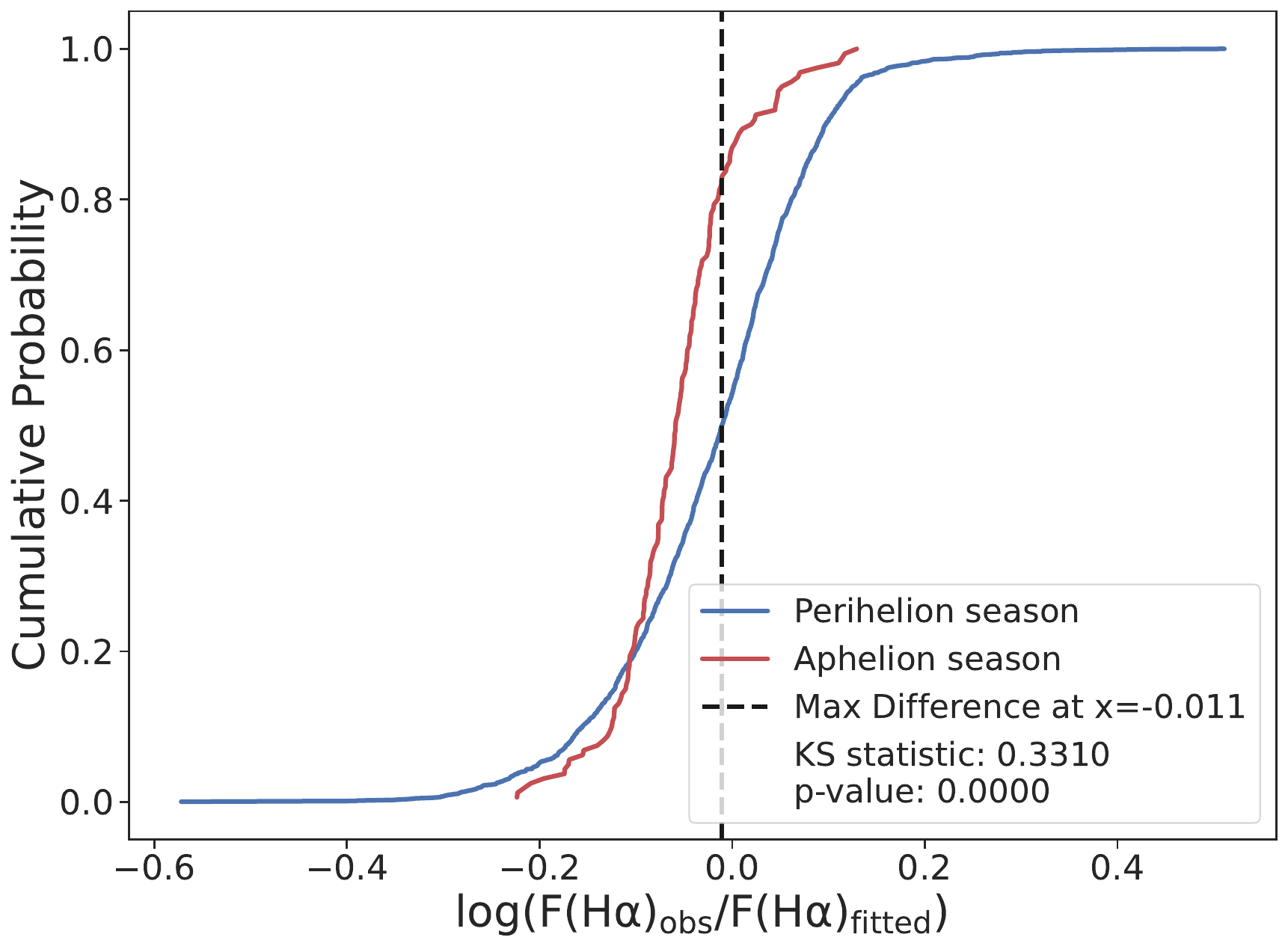}}
    \caption{Upper panel: Normalised distribution of the fitting residual in log space. Lower panel: Cumulative distribution of the fitting residual in log space. The blue and red solid lines represent data obtained near perihelion (December, January, and February) and aphelion (June, July, and August), respectively. Other symbols are the same as those in Fig. \ref{fig:geoHa_solar_cycle}.}
    \label{fig:geoHa_season}
\end{center}
\end{figure}

We selected observations during December, January, and February to be the perihelion ones (2089 exposures in total), and observations during June, July, and August to be the aphelion ones (160 exposures in total, there are few exposures in July and no exposures in August due to the monsoon season at Apache Point Observatory). There is a detectable difference between the two datasets, with the perihelion observations having a stronger observed geocoronal \Ha\ flux as expected. A simple K-S test shows that the probability that the two observation sets come from the same parent distribution is almost zero. The result is shown in Fig. \ref{fig:geoHa_season}.

\subsubsection{Intranight variation}
\begin{figure}[ht!]
    \begin{center}
    \includegraphics[width=0.90\linewidth]{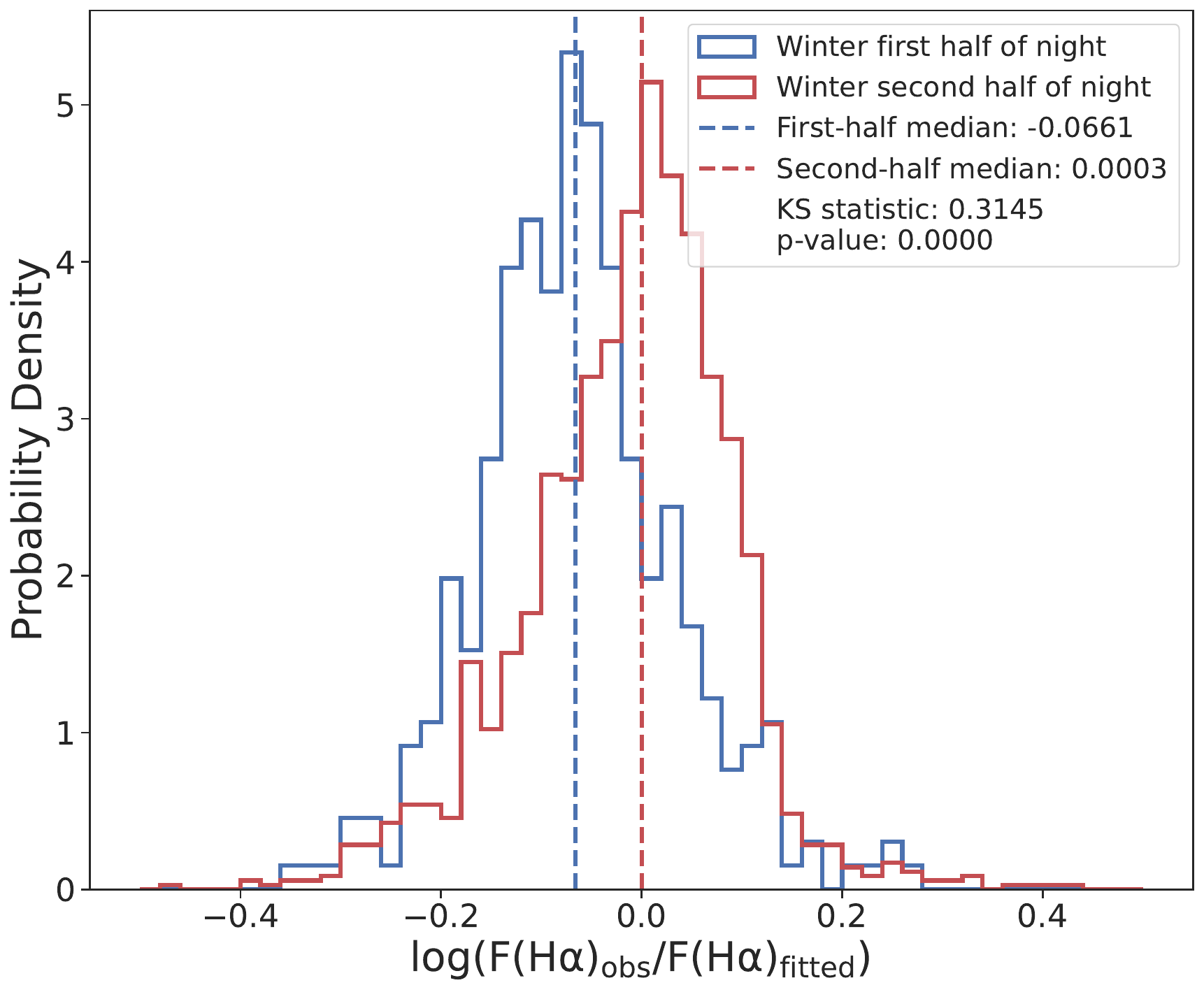}
    \caption{Normalised distribution of fitting residual in log space. Blue solid line and red solid line represent data obtained at the pre-midnight evening, and post-midnight morning during the perihelion (i.e., December, January, and February). Dashed lines represent the median value of the fitting residual in log space.}
    \label{fig:residual_vs_night_time}
    \end{center}
\end{figure}

Geocoronal \Ha\ varies meaningfully within a night. Observations from \citet{Gallant2019AM_PM_variation} suggest that geocoronal \Ha\ in the second half of the night is $\sim$30\% stronger than in the first half near the winter solstice, $\sim$6\% stronger near the summer solstice, and $\sim$15\% stronger at the equinoxes. This is associated with variations in thermospheric temperature throughout the night, which cause changes in the hydrogen density profile. We also test this using our winter observations (December, January, and February) and find a $\sim$16.5\% stronger median geocoronal \Ha\ in the second half of the night. Again, a simple K-S test shows that the probability that the two observation sets come from the same parent distribution is almost zero. The result is shown in Fig. \ref{fig:residual_vs_night_time}. We did not study this effect for summer (June, July, and August) because there are only three observations in the second half of the night during summer.

\subsubsection{Sky background reconstruction}
Imperfect sky background template reconstruction may contribute to the intrinsic scatter, as detailed in Sects.  \ref{sec:skybg_model} and \ref{sec:spec_fitting}.

The majority of the exposures have a lunar reflection of sunlight as the main factor in the modification of the solar spectrum continuum. However, atmospheric scattering and refraction would also modify the solar spectrum continuum. Since these secondary effects would have a minimum impact on the refitting procedure for correcting the solar spectrum continuum (see Eq. \ref{eq:huber_refit}), the reconstruction of the modified solar spectrum continuum would be flawed when these secondary effects become dominant.

Tropospheric scattering could also affect the sky background and strength of geocoronal \Ha\ by about 13\% to 18\%, as suggested by \citet{Geocoronal_Halpha_Fabry-Perot_facility}.

These issues have been mitigated by the use of high transparency and pre-dusk selection. However, it is noted that a more sophisticated sky background template can be used in the future.

\subsubsection{DIG contamination}
DIG contamination can contribute to the intrinsic scatter. We made the simple assumption that the pointings in the low Galactic \Ha\ group (see Sect.  \ref{subsec:data_selection}) have no Galactic \Ha\ contribution. Because we stacked the spectra for higher S/N, we have an extremely low detection limit on the order of ${\rm 10^{-20}\,erg\,cm^{-2}\,s^{-1}\,arcsec^{-2}}$, depending on the number of fibres in the IFU. Therefore, DIG emission is inevitably detected in our stacked spectrum. This DIG contamination is difficult to separate even with a significant VLSR, given the low spectral resolution and freedom of the \Ha\ absorption feature in the solar spectrum template. Fortunately, our fitted results for \Ha\ flux show a typical flux about one order of magnitude larger than the surface brightness threshold (${\rm 10^{-17.5}\ erg\ cm^{-2}\ s^{-1}\ arcsec^{-2}}$) we chose for the low Galactic \Ha\ group, and we do not find any correlation between the Galactic \Ha\ flux provided by the Finkbeiner full sky \Ha\ map and our fitted geocoronal \Ha\ flux. The exact fraction of DIG contamination depends on the shadow altitude, the Galactic altitude, and other parameters discussed above.

We note that taking the Finkbeiner Galactic \Ha\ map as a reference will inevitably introduce some DIG contamination to the low Galactic \Ha\ pointings because of the coarse spatial resolution of the map. A possible alternative would be to restrict the calibration to well-studied low-emission regions such as the Lockman Window, where \citet{Hausen2002Lockman_Window} measured only weak interstellar \Ha\ emission. However, the number of usable MaStar exposures in such restricted regions is too small to calibrate the shadow-altitude relation over a broad range of observing geometries. We therefore test a lower \Ha\ surface brightness threshold of ${\rm 10^{-18.0}\,erg\,cm^{-2}\,s^{-1}\,arcsec^{-2}}$, which selects clean sky with even less DIG contamination. The more restricted sample yields a slightly lower measurement of geocoronal F(Ha) than the original sample, indicating that there is indeed DIG contamination in the original sample. The median difference in ${\rm F(H\alpha)_{obs}}$ between the original and strict samples is ${\rm -0.11\pm^{0.03}_{0.02}\,erg\,cm^{-2}\, s^{-1}\,arcsec^{-2}}$ (the median and the 68\% uncertainty interval are both estimated from bootstrapping). We note that the DIG contamination should be independent of the shadow altitude, thereby creating a constant offset in our fitting; one may use the above difference in median ${\rm F(H\alpha)_{obs}}$ as an estimate of DIG contamination.

\subsubsection{Limitations of the empirical model}
Using a viewing geometry such as the shadow altitude alone to predict the level of geocoronal \Ha\ is insufficient. A more sophisticated and physically motivated model (e.g. the radiative transfer code \texttt{lyao\_rt} from \citealt{Bishop1999lyeo_rt}) that incorporates the Lyman $\beta$ flux and atmospheric hydrogen density and temperature profiles is needed to predict geocoronal \Ha\ more accurately. We emphasise the simplicity of the shadow altitude: it requires only the RA, Dec, and time of observation, making it suitable for any observation programme. In contrast, a radiative transfer code requires Lyman $\beta$ observations and atmospheric modelling, and geocoronal \Ha\ is very sensitive to the assumptions of the atmospheric model \citep{Bishop2001geocoronal_Ha}.

In the future, more sophisticated techniques such as PCA, or radiative transfer code can help investigate the sources of intrinsic scatter of the fitting residual between ${\rm F(H\alpha)_{obs}}$ and shadow altitude. High spatial resolution observation of the Galactic \Ha\ could also enable pointings with little to no DIG contamination. Data with minimum sky background contamination and higher spectral resolution can also be used to alleviate the problem of imperfect sky background reconstruction and DIG contamination.

\subsection{Future application/outlook} \label{subsec:future_applications}
Any intermediate spectral resolution observations around the \Ha\ window that run into the problem of removing geocoronal \Ha\ contamination will benefit greatly from this work, especially when the sky background is significant, or the observed targets are extended on the sky.

A similar approach can be done on higher-order hydrogen recombination lines, albeit the lower S/N from these recombination lines will make this more challenging.

MaStar exposures from the high Galactic \Ha\ group can also be corrected for geocoronal \Ha\ contamination using the calculated shadow altitude and the fitted parameters from Table \ref{table:shadow_alt_fit}. This will provide us with high spatial resolution and wide wavelength coverage observation of the Milky Way's interstellar medium, enabling us to study the spatial distribution, line ratios, and even velocity profiles of \hii\ regions and DIG.

Future high spatial and spectral resolution integral field spectroscopy surveys such as Affordable Multiple Aperture Spectroscopy Explorer AMASE \citep{Yan_2020, Yan2024} and Local Volume Mapper LVM from SDSS-V \citep{2019BAAS...51g.274K, drory2024sdss} also targeting the ionised gas in the Milky Way can be used. Geocoronal-\Ha-subtracted MaStar observations from the high Galactic \Ha\ group can give us a glimpse of the expected results from AMASE and LVM.

\section{Conclusion} \label{sec:conclusion}
We have demonstrated the ability to remove geocoronal \Ha\ emission from intermediate-resolution spectra from the MaStar background fibres using a simplistic sky background model and the shadow altitude. By selecting pointings with low expected Galactic \Ha\ emission, we are able to characterise the relationship between shadow altitude and geocoronal \Ha\ flux using a two-segment broken linear fit in the log-log space. The fitted result is tabulated in Table \ref{table:shadow_alt_fit}.

The root mean square of the fractional uncertainty, $\delta$, a proxy for the predictive power, is 24.41\% for our shadow altitude method, which shows a significant improvement when compared to the previous method from \citet{Zhang_2021}.

We investigated several potential sources of intrinsic scatter, including geometric factors from the LOS, solar activity, orbital distance, and an imperfect sky background model. We found tentative evidence that geocoronal emission depends on the level of solar activity, the distance between the Earth and the Sun, and the time of night. DIG contamination may also be an important contribution to the intrinsic scatter.

Future work can improve the modelling of the sky background and further investigate sources of intrinsic scatter. Overall, this technique provides reliable subtraction of geocoronal \Ha\ and enables Galactic \Ha\ studies at intermediate spectral resolution surveys.

\begin{acknowledgements}
We would like to thank Dr. Sabyasachi Chattopadhyay at SAAO for helpful suggestions on this work. We acknowledge the financial support by grants from the National Natural Science Foundation of China (No. 12425302 and 12373008) and Research Grant Council of Hong Kong (Project No. 14302522 and 14303123). RY acknowledges support by the Hong Kong Global STEM Scholar Scheme (GSP028). The research work described in this paper was conducted in the JC STEM Lab of Astronomical Instrumentation funded by The Hong Kong Jockey Club Charities Trust.

Funding for the Sloan Digital Sky Survey IV has been provided by the Alfred P. Sloan Foundation, the U.S. Department of Energy Office of Science, and the Participating Institutions. SDSS acknowledges support and resources from the Center for High-Performance Computing at the University of Utah. The SDSS website is www.sdss4.org.

SDSS is managed by the Astrophysical Research Consortium for the Participating Institutions of the SDSS Collaboration including the Brazilian Participation Group, the Carnegie Institution for Science, Carnegie Mellon University, Center for Astrophysics | Harvard \& Smithsonian (CfA), the Chilean Participation Group, the French Participation Group, Instituto de Astrofísica de Canarias, The Johns Hopkins University, Kavli Institute for the Physics and Mathematics of the Universe (IPMU) / University of Tokyo, the Korean Participation Group, Lawrence Berkeley National Laboratory, Leibniz Institut für Astrophysik Potsdam (AIP), Max-Planck-Institut für Astronomie (MPIA Heidelberg), Max-Planck-Institut für Astrophysik (MPA Garching), Max-Planck-Institut für Extraterrestrische Physik (MPE), National Astronomical Observatories of China, New Mexico State University, New York University, University of Notre Dame, Observatório Nacional / MCTI, The Ohio State University, Pennsylvania State University, Shanghai Astronomical Observatory, United Kingdom Participation Group, Universidad Nacional Autónoma de México, University of Arizona, University of Colorado Boulder, University of Oxford, University of Portsmouth, University of Utah, University of Virginia, University of Washington, University of Wisconsin, Vanderbilt University, and Yale University.
\end{acknowledgements}

\bibliographystyle{bibtex/aa}
\bibliography{bibtex/main}

\begin{appendix}

\onecolumn
\nolinenumbers
\section{Motivation of the predictive power, $\delta$}
\label{sec:delta_motivation}
\par\medskip
The residual of ${\rm F(H\alpha)_{obs}}$ from ${\rm F(H\alpha)_{fitted}}$ should increase as ${\rm F(H\alpha)_{obs}}$ increases, hence a fractional residual from ${\rm F(H\alpha)_{fitted}}$ is better to determine the predictive power of the fit. Because each data point of ${\rm F(H\alpha)_{obs}}$ yields a fractional residual and the root mean square of all fractional residual of each data point will yield $\delta$. This value is irrespective of the independent variable, allowing us to compare different methods.

\subsection{Derivation of $\delta$ in the shadow altitude method}
\label{subsec:err_analysis_shadow_alt}
The shadow altitude method evaluates the ${\rm log_{10}(F(H\alpha)_{obs})}$ as a function of ${\rm log_{10}(shadow\ altitude)}$. Since the shadow altitude is calculated from time and celestial coordinates, the uncertainty of the shadow altitude is negligible.

The residual from the fitting is in log space. To convert it into linear space, we can assume a Gaussian distribution of the residual of ${\rm log_{10}(F(H\alpha)_{obs})}$ centred at ${\rm log_{10}(F(H\alpha)_{fitted})}$, as follows:

\begin{equation}
\begin{aligned}
     \frac{\Delta {\rm F(H\alpha)}}{\rm F(H\alpha)_{fitted}} \approx {\rm log_{10}(\frac{F(H\alpha)_{obs}}{F(H\alpha)_{fitted}})*ln10},
\end{aligned}
\end{equation}
where $\Delta {\rm F(H\alpha) = F(H\alpha)_{obs} - F(H\alpha)_{fitted}}$, which is assumed to be much smaller than ${\rm F(H\alpha)_{fitted}}$. ${\rm F(H\alpha)_{fitted}}$ is evaluated at the shadow altitude of the data point using Table \ref{table:shadow_alt_fit}, ${\rm log_{10}(F(H\alpha)_{obs}/F(H\alpha)_{fitted})}$ is the residual in log space from the shadow altitude method, then $\delta$ is simply given as,

\begin{equation} \label{eq:delta_appendix}
\begin{aligned}
     \delta &= {\rm rms}(\frac{\Delta {\rm F(H\alpha)}}{\rm F(H\alpha)_{fitted}}).
\end{aligned}
\end{equation}

The Gaussian assumption is only strictly true in linear space. However, since the residual is rather small in our case here, it is justifiable to approximate a Gaussian distribution even in log space. The approximated value of $\delta$ is still useful in the comparison to the solar altitude method.

\subsection{Derivation of $\delta$ in the solar altitude method}
\label{subsec:err_analysis_solar_alt}
The solar altitude method evaluates the flux ratio between observed geocoronal \Ha\ flux and skyline ${\rm OH\lambda 6554}$, $r_{\rm obs}\equiv {\rm  F(H\alpha)_{obs}/F(OH\lambda 6554)_{obs}}$, as a function of the solar altitude. Again, the solar altitude is calculated from the time of observation, hence the residual of solar altitude is negligible.

The residual from the solar altitude method is in a flux ratio, $r$. To convert this deviation back to linear space of geocoronal \Ha\ flux, we can first find the expected geocoronal \Ha\ flux, ${\rm F(H\alpha)_{fitted}}$, as given below,

\begin{equation}
\begin{aligned}
     {\rm F(H\alpha)_{fitted}} = r_{\rm fitted}*{\rm F(OH\lambda 6554})_{obs},
\end{aligned}
\end{equation}
where ${r_{\rm fitted}}$ is evaluated at the solar altitude of the data point using Table \ref{table:solar_alt_fit}.

Now similarly assuming a Gaussian distribution of uncertainty of ${\rm r_{obs}}$ centred around ${\rm r_{fitted}}$, the fractional residual of geocoronal \Ha\ flux is simply,

\begin{equation} \label{eq:frac_uncer_Zhang}
\begin{aligned}
     \frac{\Delta {\rm F(H\alpha)}}{\rm F(H\alpha)_{fitted}} \approx\
     \pm \sqrt{(\frac{\delta {\rm F(OH\lambda 6554)_{obs}}}{\rm F(OH\lambda 6554)_{obs}})^2 + (\frac{\Delta r_{\rm fitted}}{r_{\rm fitted}})^2},
\end{aligned}
\end{equation}
where ${\rm \delta F(OH\lambda 6554})_{obs}$ is the uncertainty of ${\rm F(OH\lambda 6554)_{obs}}$ given by \texttt{scipy.optimize.curve\_fit} when fitting using Equations \ref{eq:skybg_model} and \ref{eq:solar_template_corr}, $\Delta r_{\rm fitted} = r_{\rm obs} - r_{\rm fitted}$, and the sign of this equation depends on the sign of $\Delta r_{\rm fitted}$. $\delta$ for the solar altitude method will be given by the same Eq. \ref{eq:delta_appendix} as above, therefore, the sign of Eq. \ref{eq:frac_uncer_Zhang} is irrelevant when getting the root mean square. Although the Gaussian assumption is not strictly true for the flux ratio, the approximation is justified given the small residual.

\FloatBarrier
\clearpage
\nolinenumbers
\section{Removal of weak skylines}
\label{sec:remove_weak_skylines}
\begin{figure}[htbp]
    \centering
    \includegraphics[width=0.46\textwidth]{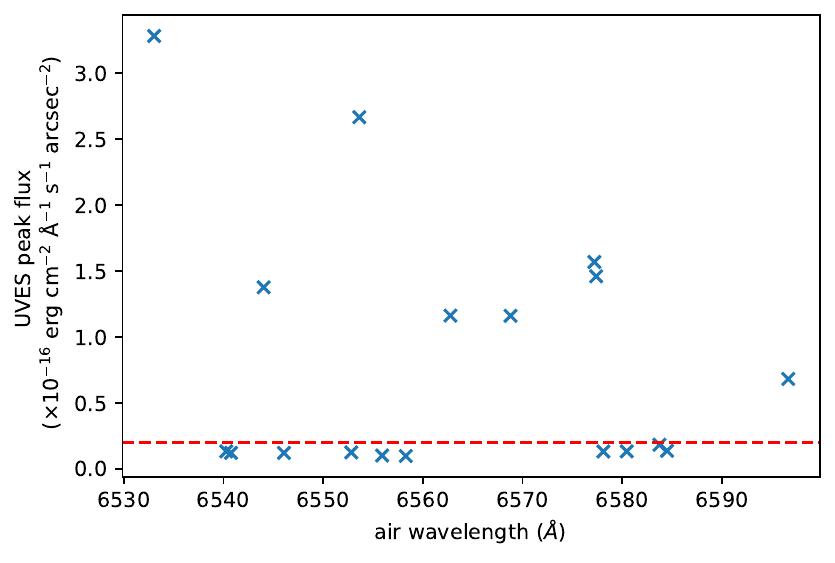}
    \hfill
    \includegraphics[width=0.46\textwidth]{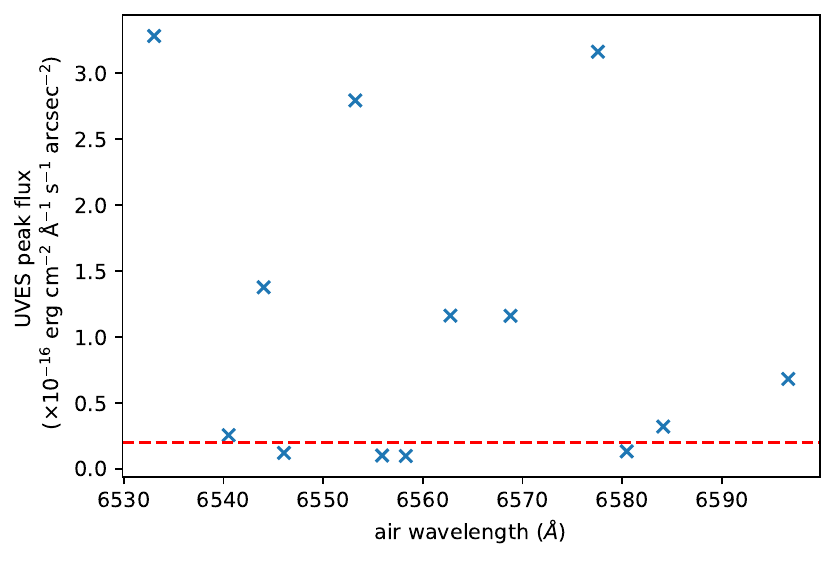}
    \caption{Left panel: UVES skyline flux before combining neighbouring skylines. Right panel: UVES skyline flux after combining neighbouring skylines. Before combination, the UVES skyline flux has a group of relatively weak skylines separated at a flux level of $2\times 10^{-17}\,\mathrm{erg\,cm^{-2}\,s^{-1}\,arcsec^{-2}}\,\text{\AA}^{-1}$ (red dashed line in both panels).}
    \label{fig:skyline_flux_threshold}
\end{figure}
\end{appendix}
\end{document}